\documentclass[12pt]{article}
\usepackage{amsfonts}

\newfont{\twelvemsb}{msbm10 scaled\magstep1}
\newfont{\eightmsb}{msbm8}
\newfam\msbfam
\textfont\msbfam=\twelvemsb \scriptfont\msbfam=\eightmsb
\catcode`\@=11
\def\Bbb{\ifmmode\let\next\Bbb@\else
\def\next{\errmessage{Use \string\Bbb\space only in math mode}}\fi\next}
\def\Bbb@#1{{\fam\msbfam{{#1}}}}

\begin{document}
\sloppy
\renewcommand{\thefootnote}{\fnsymbol{footnote}}
\newpage
\setcounter{page}{1} \vspace{0.7cm}
\begin{flushright}
YITP-04-58\\
April 2005
\end{flushright}
\vspace*{1cm}
\begin{center}
{\bf  From finite geometry exact quantities to (elliptic)
scattering amplitudes for spin chains: the 1/2-XYZ}\\
\vspace{1.8cm} {\large D.\ Fioravanti $^a$ and M.\ Rossi $^b$
\footnote{E-mail: df14@york.ac.uk, marco@yukawa.kyoto-u.ac.jp}}\\
\vspace{.5cm} $^a${\em Department of Mathematics, University of
York, Heslington,
York YO10 5DD, England} \\
\vspace{.3cm} $^b${\em Yukawa Institute for Theoretical Physics,
Kyoto University, Kyoto 606-8502,
Japan} \\
\end{center}
\renewcommand{\thefootnote}{\arabic{footnote}}
\setcounter{footnote}{0}
\begin{abstract}
{\noindent Initially,} we derive a nonlinear integral equation for
the vacuum counting function of the spin 1/2-XYZ chain in the {\it
disordered regime}, thus paralleling similar results by Kl\"umper
\cite{KLU}, achieved through a different technique in the {\it
antiferroelectric regime}. In terms of the counting function we
obtain the usual physical quantities, like the energy and the
transfer matrix (eigenvalues). Then, we introduce a double scaling
limit which appears to describe the sine-Gordon theory on
cylindrical geometry, so generalising famous results in the plane
by Luther \cite{LUT} and Johnson et al. \cite{JKM}. Furthermore,
after extending the nonlinear integral equation to excitations, we
derive scattering amplitudes involving solitons/antisolitons
first, and bound states later. The latter case comes out as
manifestly related to the Deformed Virasoro Algebra of Shiraishi
et al. \cite{SKAO}. Although this nonlinear integral equations
framework was contrived to deal with finite geometries, we prove
it to be effective for discovering or rediscovering S-matrices. As
a particular example, we prove that this unique model furnishes
explicitly two S-matrices, proposed respectively by Zamolodchikov
\cite{ZAMe} and Lukyanov-Mussardo-Penati \cite{LUK, MP} as
plausible scattering description of unknown integrable field
theories.
\end{abstract}
\vspace{1cm} {\noindent PACS}: 11.30-j; 02.40.-k; 03.50.-z
{\noindent {\it Keywords}}: Integrability; Conserved charges;
1/2-XYZ spin chain; Counting function; S-matrix.
\newpage

\section{Introduction}
By definition, two dimensional (quantum) integrable models possess
as many (possibly infinite) commuting independent (conserved)
quantities as the degrees of freedom. In the quantum scenario and
especially when the number of degrees of freedom is infinite,
integrability does not still guarantee the access to exact
information through a simple and standard way. Nevertheless, under
propitious circumstances simultaneous eigenvectors and
corresponding eigenvalues may be obtained by using various
methods, among which the Algebraic Bethe Ansatz is a very powerful
and customary technique. One reason of its great success may be
tracked in the eigenvector construction, though this reveals
itself rather cumbersome (but not available in many other Bethe
Ansatz versions). Another motive may be its huge range of
applicability, in particular within the spin chain world. As
well-known, this may be thought of as equivalent to classical
statistical lattice models (in two dimensions), which often
describe important generalisations of scaling quantum field
theories. Moreover, although the vocabulary from spin chains to
(2D) field theories has still missing items and conjectures in its
pages, it seems to have recently gained a new section on spin
chains hamiltonians as mixing matrices in four dimensional (super)
Yang-Mills theories (\cite{L,FK}, \cite{MZ} and the development
thereof originated).

 From the physical point of view, one of the most important effects
involves how the properties of the system -- and in particular the
eigenvalues of the commuting observables -- vary with its spatial
dimension, i.e. the number of sites in the spin chain case.
Besides the relevance of finite size effects in statistical
mechanics and condensed matter (e.g. \cite{ParisiCardy} and
references therein), the scale of coupling (and then of energy)
also seems to be tuned by the spin chain length in the
aforementioned 4D field theories of strong interaction.

As for finite size effects, the Non-Linear Integral Equation
(NLIE) description -- first introduced in \cite{KP} for the
conformal vacuum and then derived for an off-critical vacuum in
\cite{DDV1} by other ways -- turned out to be an efficient tool in
order to explore scaling properties of conserved charges. Since
\cite{DDV2}, still concerning features of the vacuum, and
\cite{FMQR}, regarding excited states, a number of articles was
devoted to the analysis of and through a NLIE and mainly follows
the route pioneered by Destri and de Vega \cite{DDV1} (cfr. the
Hungarian lectures \cite{Rav} for an overview). In this way (which
will be ours too), the NLIE stems directly from the Bethe
equations and characterises a quantum state by means of a single
integral equation in the complex plane. The NLIE has been widely
studied for integrable models described by trigonometric-type
Bethe equations: for instance, the 1/2-XXZ spin chain \cite{KBP},
the inhomogeneous 1/2-XXZ and sine-Gordon field theory (ground
state in \cite{DDV2}, excited states in \cite{FMQR}) and the
quantum (m)KdV-sG theory \cite{FR1}. Here instead we wish to
understand better a less studied, but more general set-up: the
elliptic one. In particular, we choose the 1/2-XYZ spin chain as
prototype of our investigation, since it is the direct
generalization of the basic trigonometric models and yet not so
much complicated to prevent a detailed and profundus analysis. Its
hamiltonian may also have serious chances to represent a mixing
matrix in some gauge theory.

In addition, we have at least another motivation to study {\it
elliptic theories} and especially elliptic spin chains. Again this
comes out partially from 2D integrable field theories. The latter
are often studied by starting from a scattering S-matrix, which
replaces somehow the Lagrangian as definition of the theory on the
plane (the most significant example being the sine-Gordon study
pioneered by both Zamolodchikov \cite{ZAM}). Very often both
S-matrix and Lagrangian (or the specific perturbation of a
conformal field theory) are well identified and tied together.
Therefore, the S-matrix has naturally become {\it the definition}
of a field theory, provided it verifies all the field theory
axioms and integrability. Nevertheless, the field theory
counterpart of a specific S-matrix is sometimes not clear in the
Lagrangian language and the assumption of the S-matrix as starting
point should be only a useful working hypothesis towards further
investigations and identifications. The elliptic case is indeed a
good example of this phenomenon with its beginning \cite{ZAMe}
coeval with the trigonometric relative \cite{ZAM}, which is
crystal clear and paradigmatic since then. On the contrary, the
whole elliptic scenario is still controversial and very subtle,
although some proposals were recently supported \cite{LS}. In this
article we want to modify the current perspective on the
correspondence problem and prove that the $8$-vertex Hamiltonian
is responsible for {\it all} the known elliptic S-matrices and
gives naturally a unitary explanation of their appearance: this
effort is willing to give a global view on a fragmented matter. Of
course, we might also interpret this as a step towards a field
theory description or as an interesting tool to generate elliptic
S-matrices (e.g. by raising up the chain spin). In any case,
scattering amplitudes may be naturally obtained from the finite
size set-up, even though these are clearly defined when the volume
is infinite. Moreover, as first pointed out in \cite{FMQR}, their
derivation requires some information about the excited states.
More specifically, if the connection with the kernel of the NLIE
may appear a promising gloss in \cite{FMQR} and a comprehensive
elaboration in \cite{BOL} as regards the sine-Gordon field theory,
it can be regarded here as a very profitable, predictive and
general tool of investigation and analysis.

Concerning the content and organisation of this article, we
address the problem of writing a NLIE satisfied by the counting
function of the spin 1/2-XYZ chain in the {\it disordered regime},
by following the more versatile route of Destri and de Vega
\cite{DDV1,DDV2,FMQR} (Section 2 about the vacuum and beginning of
Section 6 about excited states). As for the vacuum state,
Kl\"umper wrote down a similar NLIE for the spin 1/2-XYZ chain in
the {\it antiferroelectric regime} and by means of the technique
initiated in \cite{KP}. In Section 3 exact expressions of the
eigenvalues of the transfer matrix (and in particular of the
energy) are given as nonlinear functionals of the counting
function. Both consist of two terms: a contribution proportional
to the size and a finite size correction to it. Section 4 is
devoted to the trigonometric limit towards the conformal
(massless) XXZ chain, both on the NLIE and on the transfer matrix
eigenvalues. Separately (Section 5), the (trigonometric, but
massive) double scaling limit -- which yields the sine-Gordon
field theory on a cylinder -- is performed and then compared to
the results of \cite{FR1}. In Section 6, by considering the first
excitations upon the vacuum state, we are able to derive the
scattering S-matrix of the soliton/antisoliton sector (in the
repulsive regime): this matrix turns out to have an elliptic form
and coincides with that proposed by general principles (field
theory axioms; integrability: factorisation and Yang-Baxter
relation) by A. B. Zamolodchikov some time ago \cite{ZAMe}.
Furthermore, we compute the scattering factor of the lightest
soliton-antisoliton bound state in the attractive regime: on the
contrary, this manifestly coincides with the structure function of
the Deformed Virasoro Algebra (DVA) by Shiraishi-Kubo-Awata-Odake
\cite{SKAO} and hence reformulates in XYZ variables the factor
hinted by Lukyanov \cite{LUK}. Then, we thought of this factor as
that describing the scattering of the {\it fundamental elliptic
scalar particle}, and therefore as a suitable candidate to
describe an elliptic deformation of the sinh-Gordon theory. And
indeed, after another mapping of variables, it coincides with the
starting definition adopted more recently by Mussardo and Penati
\cite{MP}, although the cumbersome necessary algebra has made this
coincidence likely unnoted (cfr. also \cite{LS}). Therefore, we
find an unified arrangement for both previously known elliptic
S-matrices, though the underlying theory is not properly a field
theory: on one side the elliptic Zamolodchikov S-matrix, on the
other the DVA or Lukyanov-Mussardo-Penati factor. Eventually, some
conclusions and many perspectives come to mind and part of both is
outlined in Section 7.

\section{The lattice theory}
The spin 1/2-XYZ model with periodic boundary conditions is a
(lattice) spin chain with hamiltonian written in terms of Pauli
matrices $\sigma^{x,y,z}$
\begin{equation}
{\cal H}=-\frac {1}{2}\sum _{n=1}^N(J_x \sigma _n^x \sigma
_{n+1}^x+J_y\sigma _n^y \sigma _{n+1}^y+J_z \sigma _n^z \sigma
_{n+1}^z) \, . \label {Hamilt}
\end{equation}
Here $N$ is the number of lattice sites and because of the
periodicity the site $N+1$ is identified with the site $1$. The
three (real) coupling constants $J_x$, $J_y$ and $J_z$ may be
reparametrised (up to an overall constant) in terms of elliptic
functions (as for the notations on elliptic functions, we refer to
\cite{GRA}). In fact, after introducing the (complex) elliptic
nome $q$ ($|q|<1$), we may define the modulus ${\bf k}$ (and the
complementary modulus ${\bf k}'={\sqrt {1-{\bf k}^2}}$) and the
associated complete elliptic integral of the first kind ${\bf K}$:
\begin{eqnarray}
{\bf k}=4q^{\frac {1}{2}}\prod _{n=1}^\infty \left ( \frac
{1+q^{2n}}{1+q^{2n-1}}\right)^4 \, , \nonumber \\
{\bf K}=\frac {\pi}{2}\prod _{n=1}^\infty \left
(\frac{1+q^{2n-1}}{1+q^{2n}}\right)^2
\left(\frac{1-q^{2n}}{1-q^{2n-1}}\right)^2  \, . \label {K}
\end{eqnarray}
We also introduce the parameter ${\bf K}^\prime$ such that the
elliptic nome reads
\begin{equation}
q=e^{-\pi \frac {{\bf K}^{\prime} }{{\bf K}}}\, . \label{q}
\end{equation}
The theta-functions of nome $q$, $\theta _{ab}\left (u; i\frac
{{\bf K}'}{{\bf K}}\right )$, $a,b=0,1$, may now be defined as
\begin{eqnarray}
\theta _{00}\left (u; i\frac {{\bf K}'}{{\bf K}}\right
)&=&(-qe^{2i\pi u};q^2)(-qe^{-2i\pi
u};q^2)(q^2;q^2)=\theta_3\left(\pi u; i\frac{{\bf
K}'}{{\bf K}}\right) \, , \nonumber \\
\theta _{01}\left (u; i\frac {{\bf K}'}{{\bf K}}\right
)&=&(qe^{2i\pi u};q^2)(qe^{-2i\pi
u};q^2)(q^2;q^2)=\theta_4\left(\pi u; i\frac{{\bf K}'}{{\bf
K}}\right) \, ,
\label {thetafunc}\\
\theta _{10}\left (u; i\frac {{\bf K}'}{{\bf K}}\right
)&=&2q^{\frac {1}{4}}\cos \pi u (-q^2e^{2i\pi
u};q^2)(-q^2e^{-2i\pi u};q^2)(q^2;q^2)=\theta_2\left(\pi u;
i\frac{{\bf K}'}{{\bf K}}\right)
\, , \nonumber \\
\theta _{11}\left (u; i\frac{{\bf K}'}{{\bf
K}}\right)&=&-2q^{\frac{1}{4}}\sin \pi u(q^2e^{2i\pi
u};q^2)(q^2e^{-2i\pi u};q^2)(q^2;q^2)=-\theta_1\left(\pi u;
i\frac{{\bf K}'}{{\bf K}}\right) \, , \nonumber
\end{eqnarray}
where we have introduced a shorthand notation for the infinite
products,
\begin{equation}
(x;a)=\prod _{s=0}^{\infty}(1-xa^s) \, , \label {infprod}
\end{equation}
and at the furthest right the expressions in terms of the usual
Jacobi theta-functions $\theta_i$, $i=1,2,3,4$. With the exception
of Section 6, we will find more convenient to use the alternative
functions:
\begin{eqnarray}
H\left (u; i\frac {{\bf K}'}{{\bf K}}\right )&=&-\theta _{11}\left
(\frac {u}{2{\bf K}};i\frac {{\bf K}'}{{\bf K}} \right ) \, ,
\quad H_1\left (u; i\frac {{\bf K}'}{{\bf K}}\right )=\theta
_{10}\left (\frac {u}{2{\bf K}};i\frac {{\bf K}'}{{\bf K}}\right )
\, ,\\
\Theta \left (u; i\frac {{\bf K}'}{{\bf K}}\right )&=&\theta
_{01}\left (\frac {u}{2{\bf K}};i\frac {{\bf K}'}{{\bf K}}\right )
\, , \quad \Theta _1\left (u; i\frac {{\bf K}'}{{\bf K}}\right
)=\theta _{00}\left (\frac {u}{2{\bf K}};i\frac {{\bf K}'}{{\bf
K}}\right ) \, .
\end{eqnarray}
The Jacobian elliptic functions are
\begin{eqnarray}
{\mbox {sn}}\left (u; i\frac {{\bf K}'}{{\bf K}}\right )&=& \frac
{1}{\sqrt {{\bf k}}}\frac {H\left (u; i\frac {{\bf K}'}{{\bf
K}}\right )} {\Theta \left (u; i\frac {{\bf K}'}{{\bf K}}\right )}
\, , \quad {\mbox {cn}}\left (u; i\frac {{\bf K}'}{{\bf K}}\right
)= {\sqrt \frac {{\bf k}'}{{\bf k}}}\frac {H_1\left (u; i\frac
{{\bf K}'}{{\bf K}}\right )} {\Theta \left (u; i\frac {{\bf
K}'}{{\bf K}}\right )}
\, , \\
{\mbox {dn}}\left (u; i\frac {{\bf K}'}{{\bf K}}\right )&=&{\sqrt
{{\bf k}'}}\frac {\Theta _1\left (u; i\frac {{\bf K}'}{{\bf
K}}\right )}{\Theta \left (u; i\frac {{\bf K}'}{{\bf K}}\right )}
\, .
\end{eqnarray}
In order to simplify notations, from now on we will omit the
dependence on the elliptic nome, when the elliptic functions have
elliptic nome $q$ (\ref {q}).

In terms of Jacobian elliptic functions the coupling constants
$J_x$, $J_y$ and $J_z$ in (\ref{Hamilt}) are parametrised as
\begin{equation}
J_x=1+{\bf k}\, {\mbox {sn}}^2 2\eta \, , \quad J_y=1-{\bf k}\,
{\mbox {sn}}^2 2\eta \, , \quad J_z={\mbox {cn}}2\eta \, {\mbox
{dn}} 2\eta \, , \label {Jpar}
\end{equation}
where $\eta$ needs to be real.
\medskip
The eigenvalues of the spin 1/2-XYZ transfer matrix were firstly
found by Baxter \cite{BAX} by using his auxiliary matrix
($Q$-operator) technique. Afterwards, in the paper \cite{TF}
Takhtadjan and Faddeev wrote down also the eigenvectors (at least
formally), making use of the Algebraic Bethe Ansatz. In the
present paper, we want to restrict our analysis within the
disordered regime, i.e.
\begin{eqnarray}
&& 0<q<1  \, , \nonumber \\
&& 0<\eta < {\bf K} \, , \label {regime}
\end{eqnarray}
which in particular entail ${\bf K}>0$ and ${\bf K}'>0$. With this
choice of parameters, we differ from \cite{KLU}, where a NLIE was
written (by means of a method far from ours, though) in the
antiferroelectric regime: $0<q<1$ and $0<i\eta<{\bf K}'$. The
Bethe equations obtained in \cite {BAX,TF} are:
\begin{equation}
\left [ \frac {H(i\alpha _j +\eta)\Theta (i\alpha _j +\eta)}
{H(i\alpha _j -\eta)\Theta (i\alpha _j -\eta)}\right]^N= -e^{-4\pi
i \nu \frac {\eta}{2{\bf K}}}\prod _{k=1}^n \frac{ H(i\alpha _j -
i\alpha _k +2\eta )\Theta (i\alpha _j - i\alpha _k +2\eta )} {
H(i\alpha _j - i\alpha _k -2\eta )\Theta (i\alpha _j - i\alpha _k
-2\eta )} \, , \label{bethe}
\end{equation}
for all $j=1,...,n$, where $\nu$ is an integer. Equations (\ref
{bethe}) are valid when
\begin{equation}
m_1\eta = 2m_2 {\bf K}  \, ,  \label {relpar}
\end{equation}
where $m_1$ and $m_2$ are integers, and when $2n=N$ (mod $m_1$).
The corresponding eigenvalues of the transfer matrix are
\begin{eqnarray}
\Lambda_N(\alpha)&=&e^{2\pi i \nu \frac {\eta}{2{\bf K}}}\Theta
(0)^NH(i\alpha +\eta)^N \Theta (i\alpha +\eta)^N\prod
_{j=1}^{n}\frac {H(i\alpha _j - i\alpha +2\eta)\Theta (i\alpha _j
-i\alpha
+2\eta)}{H(i\alpha _j-i\alpha)\Theta (i\alpha _j-i\alpha)}+\nonumber \\
\label {traeig0}  \\
&+&e^{-2\pi i \nu \frac {\eta}{2{\bf K}}}\Theta (0)^NH(i\alpha
-\eta)^N \Theta (i\alpha -\eta)^N\prod _{j=1}^{n}\frac {H(i\alpha
- i\alpha _j +2\eta)\Theta (i\alpha -i\alpha _j+2\eta)}{H(i\alpha
-i\alpha _j)\Theta (i\alpha -i\alpha _j)} \, . \nonumber
\end{eqnarray}
In writing (\ref {bethe}, \ref {traeig0}) we used notations by
\cite {TF}.

\subsection{Nonlinear integral equation for the vacuum}
We now want to study the Bethe state with lowest energy, i.e. the
vacuum. From \cite{TF,TAK} we know that such a state is given by
all real roots ($\alpha _j \in {\mathbb R}$) of the Bethe
equations enjoying these additional properties
\begin{equation}
\nu =0 \, , \quad n=\frac {N}{2} \, ,  \quad - \frac {{\bf K}'}{2}
< \alpha _j < \frac {{\bf K}'}{2} \, . \label {vacuum}
\end{equation}
Using symmetry properties of elliptic functions, we rewrite
equations (\ref {bethe}) as
\begin{equation}
\left [ \frac {H(i\alpha _j +\eta)\Theta (i\alpha _j +\eta)}
{H(-i\alpha _j +\eta)\Theta (-i\alpha _j +\eta)}\right]^N=
(-1)^{1+\frac {N}{2}}\prod _{k=1}^{N/2} \frac{ H(i\alpha _j -
i\alpha _k +2\eta )\Theta (i\alpha _j - i\alpha _k +2\eta )} {
H(-i\alpha _j + i\alpha _k +2\eta )\Theta (-i\alpha _j + i\alpha
_k +2\eta )} \, . \label{bethe2}
\end{equation}
We then define the function
\begin{equation}
\phi (x,\xi)=i \ln \frac {H(\xi -ix)\Theta (\xi -ix)}{H(\xi
+ix)\Theta (\xi +ix)} \, , \quad \xi \in {\mathbb {R}} \, , \label
{phi}
\end{equation}
which has branch points occurring at the points $x=x_{r,s}$ of the
complex plane such that
\begin{equation}
|{\mbox {Re}}x_{r,s}|=r {\bf K}'  \, , \, \, |{\mbox
{Im}}x_{r,s}|=\xi - 2 s {\bf K} \, , \quad r,s \, \in \, \,
{\mathbb Z} \, \label{sing} .
\end{equation}
Therefore $\phi (x,\xi)$ is analytic for $x$ in a strip around the
real axis defined by the condition
\begin{equation}
|{\mbox {Im}}x|<{\mbox {min}}\{ \xi , |\xi -2 {\bf K}| \} \, .
\label{regionan}
\end{equation}
Having introduced the function $\phi$ (\ref {phi}), we may write
the Bethe equations as
\begin{equation}
iN\phi (\alpha _j , \eta)=\ln (-1)^{1+\frac {N}{2}} +i \sum
_{k=1}^{N/2} \phi (\alpha _j - \alpha _k , 2 \eta) \, .
\end{equation}
We define the counting function
\begin{equation}
Z_N(x)=N\phi (x,\eta)-\sum _{k=1}^{N/2} \phi (x-\alpha _k, 2\eta)
\, , \label {count}
\end{equation}
which is analytic, as a consequence of (\ref{regionan}), in the
region (containing the real axis)
\begin{eqnarray}
&& |{\mbox {Im}}x|<\eta \, , \quad {\mbox {if}} \quad 0<\eta
<\frac
{2}{3}{\bf K} \, , \nonumber \\
&& |{\mbox {Im}}x|<2{\bf K}-2\eta \, , \quad {\mbox {if}} \quad
\frac {2}{3}{\bf K}<\eta < {\bf K} \, . \label{anaregio}
\end{eqnarray}
In terms of the counting function, the Bethe roots are identified
by the condition
\begin{equation}
Z_N(\alpha _j)=\pi \left (2I_j+1+\frac {N}{2} \right ) \, , \quad
I_j \in {\mathbb Z} \, . \label{holeform}
\end{equation}
For the sake of simplicity, we restrict our analysis to the case $N\in
4 {\mathbb N}$, so that
\begin{equation}
e^{iZ_N(\alpha _j)}=-1 \, . \label{baeqsN}
\end{equation}
Now, fundamental property of the vacuum roots, $\alpha _j$, is not
to allow any {\it missing root} (hole) in equation
(\ref{holeform}) and therefore entail a simple sum property about
a function $g(x)$ analytic around the real axis
\begin{equation}
2\pi i \sum _{k=1}^{N/2}g(\alpha _k)=-\int _{-\frac {{\bf K}'}{2}
}^{\frac {{\bf K}'}{2} }dx g'(x-i\epsilon)\ln \left
[1+e^{iZ_N(x-i\epsilon)}\right] -\int _{\frac {{\bf K}'}{2}
}^{-\frac {{\bf K}'}{2} }dx g'(x+i\epsilon)\ln \left
[1+e^{iZ_N(x+i\epsilon)}\right] \, . \label {equ}
\end{equation}
In (\ref{equ}) $\epsilon>0$ is a parameter small enough to keep
the integration within the analyticity domain\footnote{We are
implicitly assuming the reasonable hypothesis that $\epsilon$ may
also be made small enough not to allow {\it spurious} solutions of
(\ref{baeqsN}) within the integration contour.} (prime means, as
usual, derivation). In the limit $\epsilon \rightarrow 0$ this
equation can be rearranged as
\begin{equation}
2\pi \sum _{k=1}^{N/2}g(\alpha _k)=-\int _{-\frac {{\bf K}'}{2}
}^{\frac {{\bf K}'}{2} }dx g'(x)Z_N(x)+2\int _{-\frac {{\bf
K}'}{2} }^{\frac {{\bf K}'}{2} }dx g'(x){\mbox {Im}}\ln \left
[1+e^{iZ_N(x+i0)}\right] \, . \label {prop}
\end{equation}
Thanks to the analyticity property (\ref{anaregio}), we may
therefore rewrite the vacuum counting function (\ref{count}) in a
useful form
\begin{equation}
Z_N(x)=N\phi (x,\eta)-\int _{-\frac {{\bf K}'}{2} }^{\frac {{\bf
K}'}{2} }\frac {dy}{2\pi} \phi '(x-y,2\eta)Z_N(y)+\int _{-\frac
{{\bf K}'}{2} }^{\frac {{\bf K}'}{2} }\frac {dy}{\pi} \phi '(x-y,
2\eta){\mbox {Im}}\ln \left [1+e^{iZ_N(y+i0)}\right] \, , \label
{ddv1}
\end{equation}
where $\phi'(x,\eta)$ is the $x$-derivative. An inspection of the
transformation properties of the elliptic functions
\begin{eqnarray}
H(u+i{\bf K}')&=&iq^{-\frac {1}{4}}e^{-\frac {i\pi u}{2\bf
K}}\Theta
(u) \, , \nonumber \\
\Theta (u+i{\bf K}')&=&iq^{-\frac {1}{4}}e^{-\frac {i\pi u}{2\bf
K}}H (u) \, , \nonumber
\end{eqnarray}
shows that $\phi (x,\xi)$ and, consequently, $Z_N(x)$ are
quasiperiodic in their regions of analyticity (\ref{regionan}) and
(\ref{anaregio}) respectively, with a real quasiperiod ${\bf K}'$:
\begin{eqnarray}
\phi (x+{\bf K}',\xi)- \phi (x,\xi)&=&2\pi \left (1-\frac
{\xi}{\bf K} \right) \,  \, \, {\mbox {($x$ in strip (\ref
{regionan})})} \, ,
\label {quaphi} \\
Z_N(x+{\bf K}')-Z_N(x)&=&\pi N \,  \, \, {\mbox {($x$ in strip
(\ref {anaregio})})} \, . \label {quaZ}
\end{eqnarray}
Obviously, the derivatives of $\phi^\prime(x,\xi)$ and
$Z^\prime_N(x)$ are periodic (with period ${\bf K}'$). A good way
to rearrange (\ref {ddv1}) is to introduce the Fourier coefficient
\begin{equation}
\hat f (n)=\int _{-\frac {{\bf K}'}{2}}^{\frac {{\bf K}'}{2}} dx
f(x) e^{2i \frac {n \pi x}{{\bf K}'}} \, , \label {fourcoe}
\end{equation}
for a (quasi)periodic function $f(x)$. In terms of the
coefficients $\hat f(n)$ the (quasi)periodic function $f(x)$ is
expressed as
\begin{equation}
f(x)=\frac {1}{{\bf K}'}\sum _{n=-\infty}^{+\infty} \hat f(n) e
^{-2i\frac {n\pi x}{{\bf K}'}} \, , \label {fourser}
\end{equation}
within the principal interval $-\frac {{\bf K}'}{2}<x <\frac {{\bf
K}'}{2}$. If we consider a periodic function $f$ and (in general)
a quasiperiodic function $g$, it is well known that their periodic
convolution,
\begin{equation}
(f\otimes g)(x)=\int _{-\frac {{\bf K}'}{2}}^{\frac {{\bf K}'}{2}}
dy f(x-y)g(y) \, ,
\end{equation}
has Fourier coefficients given by the product
\begin{equation}
\widehat {(f\otimes g)}(n)=\hat f (n) \hat g(n) \, .
\end{equation}
Bearing this property in mind, we introduce the shorter notation
\begin{equation}
L_N(x)={\mbox {Im}}\ln \left [1+e^{iZ_N(x+i0)}\right]  \, ,
\end{equation}
and may easily prove that the Fourier coefficients of all terms in
relation (\ref{ddv1}) satisfy the relation
\begin{equation}
\hat Z _N(n)=N \hat \phi (n,\eta) - \frac {1}{2\pi} \hat {\phi '}
(n, 2\eta)\hat Z _N(n) +\frac {1}{\pi} \hat {\phi '} (n, 2\eta)
\hat L_N(n) \, .
\end{equation}
And this immediately entails
\begin{equation}
\hat Z _N(n)=N \frac {\hat \phi (n,\eta) }{1+\frac {1}{2\pi} \hat
{\phi '} (n, 2\eta)}+2 \frac {\frac {1}{2\pi} \hat {\phi '} (n,
2\eta)}{1+\frac {1}{2\pi} \hat {\phi '} (n, 2\eta)}  \hat L_N(n)
\, . \label {ddv2}
\end{equation}
Being the functions $Z_N(x)$, $\phi (x)$ and $L_N(x)$ all odd,
their Fourier coefficients are vanishing for $n=0$ (equation
(\ref{ddv2}) is trivially satisfied), and hence we may drop out
the zero mode in the series expansion. Therefore for the Fourier
series we obtain
\begin{equation}
Z_N(x)=N F(x) +2 \int _{-\frac {{\bf K}'}{2}}^{\frac {{\bf
K}'}{2}}dy G(x-y) {\mbox {Im}}\ln \left [1+e^{iZ_N(y+i0)}\right]
\, , \label{ddv3}
\end{equation}
where we have defined the {\it forcing term}
\begin{equation}
F(x)=\frac {1}{{\bf K}'}\sum _{n=-\infty}^{+\infty} \frac {\hat
\phi (n,\eta) }{1+\frac {1}{2\pi} \hat {\phi '} (n, 2\eta)}e^{-2i
\frac {n \pi x}{{\bf K}'}} \, , \label {F}
\end{equation}
and the operator kernel
\begin{equation}
G(x)=\frac {1}{{\bf K}'}\sum _{\stackrel {n=-\infty}{n\not =
0}}^{+\infty}\frac {\frac {1}{2\pi} \hat {\phi '} (n,
2\eta)}{1+\frac {1}{2\pi} \hat {\phi '} (n, 2\eta)}  e^{-2i \frac
{n \pi x}{{\bf K}'}} \, , \label{G1}
\end{equation}
where we have decided to remove the zero mode.

Equation (\ref {ddv3}) is the Non-Linear Integral Equation
describing the vacuum of the spin 1/2-XYZ chain in the disordered
regime (\ref {regime}). Even though it has been derived supposing
$x$ real, it is valid in the strip (\ref{anaregio}) thanks to
analytic continuation, which would fail at some points at the
border of the strip (with real parts given by the first of
(\ref{sing})). Now, we need to compute explicitly the Fourier
series for the functions involved in (\ref {ddv3}).

\subsection{Calculation of Fourier coefficients}
First, we recall the Fourier coefficient of $\phi ^{\prime}(x,
2\eta)$ from the results of Appendix A. Since $x$ is in region
(\ref {anaregio}), we use the first of (\ref {fourphiprimegen1})
or the second of (\ref {fourphiprimegen2}) to obtain in either
case
\begin{equation}
      \hat {\phi '} (n, 2\eta)=2\pi \frac {\sinh \frac {2n({\bf
K}-2\eta)\pi}{{\bf K}'}}{\sinh \frac {2n{\bf K}\pi}{{\bf K}'}} \,
. \label {fourphiprime}
\end{equation}
Now, we are ready to compute the coefficients
\begin{equation}
\hat {\phi } (n, \eta)=\int _{-\frac {{\bf K}'}{2}}^{\frac {{\bf
K}'}{2}} dx \phi (x,\eta ) e^{2i \frac {n \pi x}{{\bf K}'}} \, .
\end{equation}
The zero mode $\hat {\phi } (0, \eta)=0$ is simply given by $\phi
(x,\eta)=-\phi (-x,\eta)$, while the others are given upon
integrating by parts and exploiting the quasi-periodicity
(\ref{quaphi}):
\begin{eqnarray}
      \hat {\phi '} (n, \eta)&=&\int _{-\frac {{\bf K}'}{2}}^{\frac {{\bf
K}'}{2}} dx \phi ^\prime (x,\eta ) e^{2i \frac {n \pi x}{{\bf
K}'}}=\nonumber \\
      &=&(-1)^n\left [\phi \left (\frac {{\bf K}'}{2}, \eta\right )-\phi
\left (-\frac {{\bf K}'}{2},\eta \right )\right ]-\int _{-\frac
{{\bf K}'}{2}}^{\frac {{\bf K}'}{2}} dx \phi (x,\eta ) \frac
{2in\pi}{{\bf
K}'}e^{2i \frac {n \pi x}{{\bf K}'}}=\nonumber \\
      &=& 2\pi \left (1-\frac {\eta}{\bf K}\right)\cos n\pi -\frac
{2in\pi}{{\bf K}'}\hat {\phi } (n, \eta) \ .
\end{eqnarray}
Plugging (\ref{fourphiprime}) into, we are given the required
expression
\begin{eqnarray}
      \hat {\phi } (0, \eta)&=&0 \, , \nonumber \\
      \hat {\phi } (n, \eta)&=&\frac {i{\bf K}'}{n}\left \{ \frac {\sinh
\frac {2n({\bf K}-\eta)\pi}{{\bf K}'}}{\sinh \frac {2n{\bf
K}\pi}{{\bf K}'}} -\left (1-\frac {\eta}{\bf K}\right)\cos n\pi
\right\},
      \, \, n\neq 0  \, ,
\label {fourphi}
\end{eqnarray}
which entails, with quasi-periodicity at hand,
\begin{equation}
\phi (x, \eta)=i \sum _{\stackrel {n=-\infty}{n \not=
0}}^{+\infty}\frac {e^{-2i \frac {n \pi x}{{\bf K}'}}}{n}\frac
{\sinh \frac {2n({\bf K}-\eta)\pi}{{\bf K}'}}{\sinh \frac {2n{\bf
K}\pi}{{\bf K}'}} +2\left (1-\frac {\eta}{\bf K}\right)\frac {\pi
x}{{\bf K}'} \,. \label {phi1}
\end{equation}
And we may re-write this series in a more compact form as
\begin{equation}
\phi (x, \eta)=\sum _{n=-\infty}^{+\infty}\frac {\sin{\frac {2n
\pi x}{{\bf K}'}}}{n}\frac {\sinh \frac {2n({\bf K}-\eta)\pi}{{\bf
K}'}}{\sinh \frac {2n{\bf K}\pi}{{\bf K}'}} \,, \label {phi1a}
\end{equation}
whose convergency domain is (like in (\ref {phi1})) $|{\mbox
{Im}}x|<\eta$, containing the strip (\ref {anaregio}).

\subsection{Forcing and kernel functions}
Given the preceding expressions of $\hat{\phi}(n,\eta)$
(\ref{fourphi}) and of $\hat{\phi '} (n, \eta)$
(\ref{fourphiprime}), we may explicitly mould the forcing term
(\ref{F}) into
\begin{equation}
F(x)=\sum _{\stackrel {n=-\infty}{n \not= 0}}^{+\infty} \frac
{i}{n}\frac {\sinh \frac {2n({\bf K}-\eta)\pi}{{\bf K}'} - \left
(1-\frac {\eta}{\bf K}\right)\cos n\pi \sinh \frac {2n{\bf
K}\pi}{{\bf K}'}}{\sinh \frac {2n{\bf K}\pi}{{\bf K}'}+\sinh \frac
{2n({\bf K}-2\eta)\pi}{{\bf K}'}}e^{-2i \frac {n \pi x}{{\bf K}'}}
\, . \label {force}
\end{equation}
     From (\ref{G1}) we read off the Fourier coefficient
\begin{equation}
\hat G(n)=\frac {\frac {1}{2\pi} \hat {\phi '} (n, 2\eta)}{1+\frac
{1}{2\pi} \hat {\phi '} (n, 2\eta)} \, ,
\end{equation}
and then simplify this by means of (\ref{fourphiprime}),
\begin{equation}
\hat G(n)=\frac {\sinh \frac {2n({\bf K}-2\eta)\pi}{{\bf
K}'}}{2\sinh \frac {2n({\bf K}-\eta)\pi}{{\bf K}'} \cosh \frac
{2n\eta \pi}{{\bf K}'}} \, , \label{hatG}
\end{equation}
whence to obtain the final expression for the kernel
\begin{equation}
G(x)=\frac {1}{{\bf K}'}\sum _{\stackrel {n=-\infty}{n \not=
0}}^{+\infty} \frac {\sinh \frac {2n({\bf K}-2\eta)\pi}{{\bf
K}'}}{2\sinh \frac {2n({\bf K}-\eta)\pi}{{\bf K}'} \cosh \frac
{2n\eta \pi}{{\bf K}'}}  e^{-2i \frac {n \pi x}{{\bf K}'}} \, .
\label {G2}
\end{equation}
\medskip
{\bf {Remark.}} In the case $\eta ={\bf K}/2$ the kernel function
$G(x)$ vanishes. Therefore the solution of the NLIE is trivial:
\begin{equation}
Z_N(x)=N \sum _{n=-\infty}^{+\infty}\frac {\sin{\frac {2n \pi
x}{{\bf K}'}}}{2n \cosh \frac {n\pi {\bf K}}{{\bf K}'}}= N {\mbox
{am}}\left(2x; i\frac {\bf {K}}{\bf {K}'}\right) = N {\mbox
{arcsin}}\,  {\mbox {sn}}\left (2x; i\frac {\bf {K}}{\bf
{K}'}\right) \, ,
\end{equation}
where the elliptic functions have nome $q' = e^{-\pi \frac {\bf
{K}}{\bf {K}'}}$. This is not a surprise as this case reduces to
the free-fermion point $\eta =\pi/4$ in the trigonometric
($q\rightarrow 0$, cfr. Section 4) limit.

\section{Eigenvalues of the transfer matrix}
\setcounter{equation}{0} It is the main aim of this Section to
write down the eigenvalue of the transfer matrix (\ref{traeig0})
on the vacuum state in terms of the solution of the (vacuum)
Non-Linear Integral Equation (\ref{ddv3}). Although this might
seem a priori a complication, it has at least two main advantages.
First, instead of solving a big number of transcendental (Bethe)
equations, one should solve a NLIE: this makes numerical
computations and approximations much simpler. Second, in the
following expressions the bulk terms (proportional to the size $N$
of the system) are clearly separated from their finite size
corrections. And as it will be clear later on, the finite size
corrections and properties can be singled out in the limit
$N\rightarrow \infty$.

Again, we start from relation (\ref {prop}) which expresses the
sum on the vacuum Bethe roots of an arbitrary analytic function
$g$. Using the NLIE, (\ref{prop}) may be written as follows
\begin{eqnarray}
&&\sum _{k=1}^{N/2}g(\alpha _k)=-\frac {N}{2\pi}\int _{-\frac
{{\bf
K}'}{2} }^{\frac {{\bf K}'}{2} }dx g'(x)F(x)+\nonumber \\
&+&\int _{-\frac {{\bf K}'}{2} }^{\frac {{\bf K}'}{2} }\frac
{dx}{\pi} g'(x)\int _{-\frac {{\bf K}'}{2} }^{\frac {{\bf K}'}{2}
}dy [\delta (x-y)-G(x-y)]{\mbox {Im}}\ln \left
[1+e^{iZ_N(y+i0)}\right] \, . \label {prop2}
\end{eqnarray}
In terms of the Fourier coefficients this relation reads as well
\begin{eqnarray}
&&\sum _{k=1}^{N/2}g(\alpha _k)=-\frac {N}{2\pi {\bf K}'}\sum
_{n=-\infty}^{+\infty} \hat F(n) \hat {g'}(-n) +\nonumber \\
&+& \int _{-\frac {{\bf K}'}{2} }^{\frac {{\bf K}'}{2} }\frac
{dx}{\pi} \left \{ \frac {1}{{\bf K}'} \sum _{n=-\infty}^{+\infty}
\hat {g'}(n)[1-\hat G(n)] e^{-\frac {2in\pi x}{{\bf K}'} } \right
\} {\mbox {Im}}\ln \left [1+e^{iZ_N(x+i0)}\right]  \, . \label
{prop3}
\end{eqnarray}
Therefore, once the solution of (\ref {ddv3}) is found, the
relation (\ref{prop3}) may be used in order to compute (very often
numerically) the eigenvalues of observables on the vacuum. We
remark that the first term in (\ref {prop3}) is apparently
proportional to $N$. Therefore it gives usually the leading order
in the $N\rightarrow \infty$ limit. In fact, the second term in
(\ref{prop3}) usually provides the subleading corrections in that
limit. It is convenient to define the function constructed upon
$g(x)$
\begin{equation}
J_g(x)=\frac {1}{{\bf K}'} \sum _{n=-\infty}^{+\infty}  \hat
{g'}(n)[1-\hat G(n)] e^{-\frac {2in\pi x}{{\bf K}'} }  \, , \label
{Jg}
\end{equation}
which appears explicitly in (\ref {prop3}). Here we decide to add
to the definition of $G(x)$ its zero mode, since its inclusion
does not change the value of the integral in (\ref {prop3}) and
makes at the same time formul{\ae} more compact. In particular, we
want to apply the result (\ref {prop3}) to the vacuum eigenvalues
of the transfer matrix (\ref {traeig0}). In this respect, we may
introduce the decomposition
\begin{eqnarray}
\Lambda_N(\alpha)&=&\Theta (0)^NH(i\alpha +\eta)^N \Theta (i\alpha
+\eta)^N\prod _{j=1}^{N/2}\frac {H(i\alpha _j - i\alpha
+2\eta)\Theta (i\alpha _j -i\alpha +2\eta)}{H(i\alpha
_j-i\alpha)\Theta (i\alpha
_j-i\alpha)}+\nonumber \\
&+&\Theta (0)^NH(i\alpha -\eta)^N \Theta (i\alpha -\eta)^N\prod
_{j=1}^{N/2}\frac {H(i\alpha  - i\alpha _j +2\eta)\Theta (i\alpha
-i\alpha _j+2\eta)}{H(i\alpha -i\alpha _j)\Theta (i\alpha -i\alpha
_j)}=\nonumber \\
&=& \Lambda _N^+(\alpha)+\Lambda _N^-(\alpha) \, . \label {traeig}
\end{eqnarray}
Let us first concentrate on $\ln \Lambda _N^+(\alpha)$. We have
that
\begin{equation}
\ln \Lambda _N^+(\alpha)=N \ln \Theta (0)+N\ln H(i\alpha
+\eta)+N\ln \Theta (i\alpha +\eta)+ \sum _{k=1}^{N/2}\gamma
_+(\alpha _k, \alpha) \, ,
\end{equation}
where
\begin{equation}
\gamma_+(x, \alpha)=\ln \frac {H(ix - i\alpha +2\eta)\Theta (ix
-i\alpha +2\eta)}{H(ix-i\alpha)\Theta (ix-i\alpha)} \, . \label
{g+}
\end{equation}
Because of the periodicity property
\begin{equation}
\ln \Lambda _N^+(\alpha+2i{\bf K})=\ln \Lambda _N^+(\alpha) \, ,
\label {gperio}
\end{equation}
we can restrict the complex $\alpha$ to the strip
\begin{equation}
-2\eta <{\mbox {Im}}\alpha <2{\bf K}-2\eta \, . \label {strip}
\end{equation}
Upon comparing (\ref {g+}) with (\ref {phi}), we see that
\begin{equation}
\gamma_+(x,\alpha)=-i \phi (x+i{\bf K}-i\eta -\alpha, {\bf
K}-\eta) \, . \label {g-phi}
\end{equation}
Recalling (\ref{fourphiprimegen}) we easily obtain the Fourier
coefficients
\begin{equation}
\hat {\gamma ^\prime}_+(n,\alpha)=\int _{-\frac {{\bf K}'}{2}
}^{\frac {{\bf K}'}{2} } dx \frac {d}{dx} \gamma_+(x,
\alpha)e^{\frac {2in\pi x}{{\bf K}'} } \, ,
\end{equation}
in a form depending on the imaginary part of $\alpha$
\begin{eqnarray}
\hat {\gamma ^\prime}_+(n,\alpha)&=&2\pi i \frac {\sinh \frac
{2n({\bf K}-\eta)\pi}{{\bf K}'}}{\sinh \frac {2n{\bf K}\pi}{{\bf
K}'}} e^{\frac {2n\pi}{{\bf K}'}(i\alpha - \eta)} \, , \quad
{\mbox {if}} \, \, -2\eta <{\mbox {Im}}\alpha <0 \, ,
\nonumber \\
\label {hatg+} \\
\hat {\gamma ^\prime}_+(n,\alpha)&=&-2\pi i \frac {\sinh \frac {2n
\eta\pi}{{\bf K}'}}{\sinh \frac {2n{\bf K}\pi}{{\bf K}'}} e^{\frac
{2n\pi}{{\bf K}'}({\bf K}+i\alpha - \eta)} \, , \quad {\mbox {if}}
\, \, 0 <{\mbox {Im}}\alpha <2{\bf K}-2\eta \, . \nonumber
\end{eqnarray}
In this case, the function $J_{\gamma _+}$ (\ref {Jg}) equals
\begin{eqnarray}
J_{\gamma _+}(x,\alpha)&=&\frac {i\pi}{{\bf K}'}\sum
_{n=-\infty}^{+\infty}\frac {e^{\frac {2in\pi}{{\bf K}'}(\alpha
+i\eta-x)}}{\cosh \frac {2n\eta \pi}{{\bf K}'}} \, , \quad {\mbox
{if}}
\, \, -2\eta <{\mbox {Im}}\alpha <0 \, , \nonumber \\
\label {Jg+}\\
J_{\gamma _+}(x,\alpha)&=&-\frac {i\pi}{{\bf K}'}\sum
_{n=-\infty}^{+\infty}\frac {\sinh \frac {2n\eta \pi}{{\bf
K}'}}{\sinh \frac {2n({\bf K}-\eta)\pi}{{\bf K}'}\cosh \frac
{2n\eta \pi}{{\bf K}'}} e^{\frac {2in\pi}{{\bf K}'}(\alpha
+i\eta-i{\bf K}-x)} \, , \quad {\mbox {if}} \, \, 0 <{\mbox
{Im}}\alpha <2{\bf K}-2\eta \, . \nonumber
\end{eqnarray}
As a consequence, from relation (\ref {prop}) and from the
expression for $\hat F(n)$ (\ref {force}) we are finally given the
two cumbersome expressions:

$\bullet$ if $ -2\eta <{\mbox {Im}}\alpha <0 $:
\begin{eqnarray}
&&\ln \Lambda _N^+(\alpha)=N \ln \Theta (0)+N\ln H(i\alpha
+\eta)+N\ln
\Theta (i\alpha +\eta)+\nonumber \\
&+& N \sum _{\stackrel {n=-\infty}{n \not= 0}}^{+\infty} \frac
{e^{-\frac {2n\pi}{{\bf K}'}(i\alpha -\eta)}}{n}\left [ \frac
{\sinh \frac {2n({\bf K}-\eta)\pi}{{\bf K}'}}{2\cosh \frac {2n\eta
\pi}{{\bf K}'}\sinh \frac {2n{\bf K}\pi}{{\bf K}'}}-\frac {\left
(1-\frac {\eta}{{\bf K}}\right)\cos n\pi }{2 \cosh \frac {2n\eta
\pi}{{\bf K}'}}
\right ] +\nonumber \\
&&\label {tra1} \\
&+& \int _{-\frac {{\bf K}'}{2} }^{\frac {{\bf K}'}{2} }dx\frac
{i}{{\bf K}'}\sum _{n=-\infty}^{+\infty}\frac {e^{\frac
{2in\pi}{{\bf K}'}(\alpha +i\eta-x)}}{\cosh \frac {2n\eta
\pi}{{\bf K}'}} {\mbox {Im}}\ln \left [1+e^{iZ_N(x+i0)}\right] \,
; \nonumber
\end{eqnarray}

$\bullet$ if $ 0 <{\mbox {Im}}\alpha <2{{\bf K}}-2\eta $:
\begin{eqnarray}
&&\ln \Lambda _N^+(\alpha)=N \ln \Theta (0)+N\ln H(i\alpha
+\eta)+N\ln
\Theta (i\alpha +\eta)-\nonumber \\
&-& N \sum _{\stackrel {n=-\infty}{n \not= 0}}^{+\infty} \frac
{e^{-\frac {2n\pi}{{\bf K}'}({{\bf K}}+i\alpha -\eta)}}{n}\left [
\frac {\sinh \frac {2n\eta\pi}{{\bf K}'}}{2\cosh \frac {2n\eta
\pi}{{\bf K}'}\sinh \frac {2n{\bf K}\pi}{{\bf K}'}}-\frac {\left
(1-\frac {\eta}{{\bf K}}\right)\cos n\pi \sinh \frac
{2n\eta\pi}{{\bf K}'}}{2 \cosh \frac {2n\eta \pi}{{\bf K}'}\sinh
\frac {2n({{\bf
K}}-\eta)\pi}{{\bf K}'}} \right ] -\nonumber \\
&&\label {tra2} \\
&-& \int _{-\frac {{\bf K}'}{2} }^{\frac {{\bf K}'}{2} }dx\frac
{i}{{\bf K}'}\sum _{n=-\infty}^{+\infty}\frac {\sinh \frac
{2n\eta\pi}{{\bf K}'}}{\sinh \frac {2n({{\bf K}}-\eta)\pi}{{\bf
K}'} \cosh \frac {2n\eta \pi}{{\bf K}'}} e^{\frac {2in\pi}{{\bf
K}'}(\alpha +i\eta -i{\bf K}-x)}{\mbox {Im}}\ln \left
[1+e^{iZ_N(x+i0)}\right] \, . \nonumber
\end{eqnarray}
\medskip
For what concerns $\ln \Lambda _N^-(\alpha)$, we have that
\begin{equation}
\ln \Lambda _N^-(\alpha)=N \ln \Theta (0)+N\ln H(i\alpha
-\eta)+N\ln \Theta (i\alpha -\eta)+ \sum _{k=1}^{N/2}\gamma
_-(\alpha _k, \alpha) \, ,
\end{equation}
where
\begin{equation}
\gamma _-(x, \alpha)=\ln \frac {H(i\alpha -ix+2\eta)\Theta
(i\alpha -ix+2\eta)}{H(i\alpha-ix)\Theta (i\alpha-ix)} \, . \label
{g-}
\end{equation}
We notice that $\gamma _-(x,\alpha)=-\gamma _+(x,\alpha -2i\eta)$.
By using this link, the expression for $\ln \Lambda _N^-(\alpha)$
may be easily obtained from (\ref{tra1}, \ref{tra2}).

\subsection{The energy}
It may be of some interest to write down the vacuum eigenvalue of
the Hamiltonian (\ref {Hamilt}) in a form which separates the term
proportional to $N$ to its finite size correction. From \cite{TF}
we learn that in general we may extract the eigenvalues of the
Hamiltonian (of the spin 1/2-XYZ chain) from the eigenvalues of
the transfer matrix as
\begin{equation}
E_N=i{\mbox {sn}} 2 \eta \left . \frac {d}{d\alpha} \ln \Lambda_N
(\alpha) \right |_{\alpha =-i\eta} \, , \label {energy}
\end{equation}
which in the notations of last subsection takes on the form
\begin{equation}
E_N=i{\mbox {sn}} 2 \eta \left . \frac {d}{d\alpha} \ln \Lambda_N
^+(\alpha) \right |_{\alpha =-i\eta} \, .
\end{equation}
In the particular case of the vacuum, result (\ref{tra1}) yields
the wanted expression
\begin{eqnarray}
E_N^{(vac)}&=&-N {\mbox {sn}} 2 \eta \left [ \frac
{H'(2\eta)}{H(2\eta)}+ \frac {\Theta '(2\eta)}{\Theta (2\eta)}
\right
]+\nonumber \\
&+&i {\mbox {sn}} 2 \eta \Bigl \{ -N \sum _{\stackrel
{n=-\infty}{n \not = 0}}^{+\infty}\frac {2\pi i}{{\bf K}'}\left
[ \frac {\sinh \frac {2n({\bf K}-\eta)\pi}{{\bf K}'}}{2\cosh \frac
{2n\eta \pi}{{\bf K}'} \sinh \frac {2n{\bf K}\pi}{{\bf K}'}}-\frac
{\left (1-\frac {\eta}{{\bf K}}\right )\cos n\pi }{2\cosh \frac
{2n\eta \pi}{{\bf K}'}}\right]+
\label {eigham} \\
&+&\int _{-\frac {{\bf K}'}{2} }^{\frac {{\bf K}'}{2} } dx \frac
{i}{{\bf K}'}\sum _{n=-\infty }^{+\infty} \frac {2in\pi}{{\bf
K}'}\frac {e^{-\frac {2in\pi x}{{\bf K}'}}}{\cosh \frac {2n\eta
\pi}{{\bf K}'}} {\mbox {Im}}\ln \left [1+e^{iZ_N(x+i0)}\right]
\Bigr \} \, . \nonumber
\end{eqnarray}
As easily follows from this explicit expression, the last term
gives the finite size corrections to the first two terms in the
limit $N\rightarrow \infty$.

\section{Trigonometric limit (i.e. XXZ chain)}
\setcounter{equation}{0} In order to have a check on the validity
of our results and to prepare the ground for the more refined
limit of next Section, it is important to explore the
trigonometric limit in which the spin 1/2-XYZ chain reduces to the
spin 1/2-XXZ chain ($q\rightarrow 0$ or $J_x=J_y$ in (\ref
{Hamilt})). We expect to reproduce the results of papers
\cite{KBP,DDV1}. As the trigonometric limit is also expressed by
the limit ${\bf K}^\prime  \rightarrow +\infty$ (therefore ${\bf
K} \rightarrow \pi/2$), a Fourier sum can be replaced by an
integral, according to the prescription
\begin{equation}
\frac {1}{{\bf K}'}\sum _{n=-\infty}^{+\infty}  f\left (\frac
{n}{{\bf K}'}\right) \rightarrow \int _{-\infty}^{+\infty} dp f(p)
\, . \label{presc}
\end{equation}
In this Section we aim at writing down the trigonometric limit of
the vacuum NLIE (\ref {ddv3}) and of the vacuum eigenvalues of the
transfer matrix (\ref {tra1}, \ref {tra2}) and of the energy (\ref
{eigham}).  With reference to (\ref {regime}), the range of
variation of $\eta$ becomes $0<\eta<\pi/2$. By applying
prescription (\ref{presc}) to (\ref{G2}), we obtain an integral
expression for the kernel function $G(x)$
\begin{eqnarray}
G(x)\rightarrow G_{sG}(x)&=&\int_{-\infty}^{+\infty} dp \, e^{-2i
p \pi x} \, \frac {\sinh \left [ 2p\left (\frac
{\pi}{2}-2\eta\right )\pi\right ]}{2\sinh \left [2p\left (\frac
{\pi}{2}-\eta\right)\pi
\right ] \cosh 2p\eta \pi }  =\nonumber \\
&=& \int_{-\infty}^{+\infty} \frac {dp}{2\pi} \, e^{i p  x} \,
\frac {\sinh p\left (\frac {\pi}{2}-2\eta\right )}{2\sinh p\left
(\frac {\pi}{2}-\eta\right) \, \cosh p\eta  } \, . \label {G3}
\end{eqnarray}
As a check we may notice that the inclusion of the zero mode of
$G$ is irrelevant because
\begin{equation}
\frac {1}{{\bf K}^\prime}\hat G(0)=\frac {1}{2{\bf K}^\prime}\frac
{{\bf K}-2\eta}{{\bf
K}-\eta}
\end{equation}
is vanishing in that limit. Analogously the trigonometric limit of
\begin{equation}
\phi ^{\prime}(x, 2\eta)=\frac {1}{{\bf K}'}\sum
_{n=-\infty}^{+\infty}e^{-2i \frac {n \pi x}{{\bf K}'}}\hat \phi
^{\prime}(n, 2\eta)
\end{equation}
may be easily computed starting from (\ref {fourphiprime}):
\begin{equation}
\phi ^{\prime}(x, 2\eta) \rightarrow  \int
_{-\infty}^{+\infty}dp\,  e^{-i p  x} \, \frac {\sinh p\left
(\frac {\pi}{2}-2\eta \right)}{\sinh p\frac {\pi }{2}} \, . \label
{F2}
\end{equation}
For what concerns the forcing term $F(x)$, in the trigonometric
limit  the oscillating term in (\ref{force}) gives no contribution
and leaves us with
\begin{equation}
F(x)\rightarrow \frac{1}{2}\int_{-\infty}^{+\infty}\frac
{dp}{p}\frac {\sin px}{\cosh p \eta} =\frac {1}{i}\ln \tan \left
(\frac {i\pi x}{4\eta}+\frac {\pi}{4}\right) ={\mbox {arctan}}
\sinh \frac {\pi x}{2\eta} \, . \label{forcetri}
\end{equation}
Therefore, in the trigonometric limit the NLIE (\ref {ddv3}) takes
on the form
\begin{equation}
Z_N(x)=N{\mbox {arctan}} \sinh \frac {\pi x}{2\eta} +2\int
_{-\infty}^{+\infty}dy G_{sG}(x-y){\mbox {Im}}\ln \left
[1+e^{iZ_N(y+i0)}\right] \, , \label {ddvtrig}
\end{equation}
which is the Non-Linear Integral Equation for the spin 1/2-XXZ
model in the  massless antiferromagnetic regime (i.e. $J_x=J_y=1,
J_z= \cos 2 \eta$). Moreover, we may straightforwardly perform the
limit on the vacuum eigenvalue of the transfer matrix, or better
on (\ref {tra1}, \ref {tra2}), with the outcome
\medskip

$\bullet$ if $ -2\eta <{\mbox {Im}}\alpha <0 $:
\begin{eqnarray}
&&\ln \Lambda _N^+(\alpha)= N \int _{-\infty }^{+\infty }\frac
{dp}{p}e^{-ip\alpha+p\eta}\frac {\sinh p \left (\frac
{\pi}{2}-\eta\right)}{2\cosh p\eta \sinh \frac {p \pi}{2}}+\nonumber \\
&&\label {tra3} \\
&+&\int _{-\infty }^{+\infty }dx \frac {1}{2\eta \, \sinh \frac
{\pi}{2 \eta}(\alpha -x) } {\mbox {Im}}\ln \left
[1+e^{iZ_N(x+i0)}\right]  \, ; \nonumber
\end{eqnarray}

$\bullet$ if $ 0 <{\mbox {Im}}\alpha <\pi-2\eta $:
\begin{eqnarray}
&&\ln \Lambda _N^+(\alpha)=- N \int _{-\infty }^{+\infty }\frac
{dp}{p}e^{-ip\alpha+p\eta-p\frac {\pi}{2}}
\frac {\sinh p \eta }{2\cosh p\eta \sinh \frac {p \pi}{2}}- \nonumber \\
&&\label {tra34} \\
&-&\int _{-\infty }^{+\infty }dx \int _{-\infty }^{+\infty }\frac
{dp}{2\pi} \frac {i \sinh p \eta }{\sinh p \left (\frac
{\pi}{2}-\eta\right) \cosh p\eta} e^{ip\left (\alpha +i\eta-i\frac
{\pi}{2}-x\right)}{\mbox {Im}}\ln \left [1+e^{iZ_N(x+i0)}\right] .
\nonumber
\end{eqnarray}
\medskip
Eventually, the energy (\ref {eigham}) takes on the limit
expression
\begin{eqnarray}
E&\rightarrow& - N \cos 2 \eta +\pi N \sin 2 \eta  \int _{-\infty
}^{+\infty } dp \frac {\sinh 2p \left (\frac
{\pi}{2}-\eta\right)\pi
}{\sinh p\pi ^2 \cosh 2p\eta \pi} - \nonumber \\
&& \label {En} \\
&-&\frac {\pi}{4 \eta ^2}\sin 2 \eta \int _{-\infty }^{+\infty }dx
\frac {\sinh \frac {\pi x}{2\eta}}{\cosh ^2 \frac {\pi
x}{2\eta}}{\mbox {Im}}\ln \left [1+e^{iZ_N(x+i0)}\right] \, .
\nonumber
\end{eqnarray}
And indeed these formul{\ae} coincide with those given in
\cite{KBP,DDV1} as for the spin 1/2-XXZ model.

\section{A double scaling limit: cylinder sine-Gordon}
\setcounter{equation}{0} In this Section we want to rewrite the
NLIE (\ref{ddv3}) in a particular limit such that it will describe
the sine-Gordon field theory on a cylinder. This procedure extends
the celebrated results of \cite{LUT} and \cite{JKM} which concern
how a peculiar infinite length \footnote{The length is defined as
$R=N\Delta$, with the lattice spacing $\Delta$.} limit of the XYZ
reproduces sine-Gordon on the plane. We first notice that
expression (\ref {force}) for the forcing term may be separated as
\begin{equation}
NF(x)=N\sum _{\stackrel {n=-\infty}{n \not= 0}}^{+\infty} \frac
{i}{2n}\frac {e^{-2i \frac {n \pi x}{{\bf K}'}}}{\cosh \frac {2n
\eta \pi}{{\bf K}'}} -N \sum _{\stackrel {n=-\infty}{n \not=
0}}^{+\infty} \frac {i}{2n} \frac { \left (1-\frac {\eta}{\bf
K}\right)\sinh \frac {2n{\bf K}\pi}{{\bf K}'}}{\sinh \frac
{2n({\bf K}-\eta)\pi}{{\bf K}'}\cosh \frac {2n \eta\pi}{{\bf
K}'}}e^{-2i \frac {n \pi }{{\bf K}'}\left (x-\frac {1}{2}{\bf K}'
\right )} \, . \label {force1}
\end{equation}
Then, we shift the variables $x$ and $y$ in (\ref {ddv3}) as
$x=x'+\Theta$ and $y=y'+\Theta$, assuming $\Theta$ \, $x$-,
$y$-independent and positive:
\begin{eqnarray}
&&Z_N(x'+\Theta)=N F(x'+\Theta) +2 \int _{-\frac {{\bf
K}'}{2}-\Theta}^{\frac {{\bf K}'}{2}-\Theta}dy' G(x'-y') {\mbox
{Im}}\ln \left [1+e^{iZ_N(y'+\Theta+i0)}\right]   \, , \nonumber \\
      &&\quad -\frac {{\bf K}'}{2}-\Theta <x'<\frac {{\bf K}'}{2}-\Theta
\, . \label{ddv3'}
\end{eqnarray}
Now, the double scaling limit is realised by allowing ${\bf
K}'\rightarrow +\infty $ when $N \rightarrow \infty$ according to
\begin{equation}
{\bf K}'=d \ln DN \, , \quad   d >0 \, ,  \, D>0 \, ,
\label{property1}
\end{equation}
provided that $\Theta$ also diverges as
\begin{equation}
\Theta =c \ln C N \, , \quad c >0 \, , \, c <\frac{d}{2} \, , \,
C>0 \, . \label{property2}
\end{equation}
The reason why this is a double limit (namely the lattice spacing
$\Delta$ is going to zero too, whereas the length $R=N\Delta$
remains finite) is encapsulated in the coefficients $C$ and $D$
and will be clear below. Indeed, since the shift (\ref{ddv3'}) we
had in mind to define a modified counting function too:
\begin{equation}
Z(x')=\lim _{N \rightarrow \infty} Z_N(x'+\Theta) \, .
\label{limzeta}
\end{equation}
Of course the range of variation of the new independent variables
$x'$ and $y'$ gets more involved
\begin{equation}
-\frac {d}{2} \ln DN - c \ln CN <x',y'<\frac {d}{2} \ln DN - c \ln
CN \, . \label{rangexy}
\end{equation}
Now, let us derive the limiting value of the forcing term $N
F(x'+\Theta)$: the sums in (\ref{force1}) may be replaced by
integrals according to the rule (\ref{presc}), i.e.
\begin{eqnarray}
&&NF(x'+\Theta)=\frac {iN}{2}\int _{-\infty}^{+\infty} \frac
{dp}{p}\frac {e^{-2ip\pi(x'+c \ln CN)}}{\cosh 2p\eta \pi}-\nonumber \\
&-&\frac {iN}{2}\int _{-\infty}^{+\infty} \frac {dp}{p}\frac
{\left (1-\frac {2\eta}{\pi}\right)\sinh p\pi ^2\, e^{-2ip\pi(x'+c
\ln CN-\frac {d}{2}\ln DN)}}{\sinh \left [ 2p\pi \left (\frac
{\pi}{2}-\eta\right)\right] \cosh 2p\eta \pi}.
\end{eqnarray}
Because of properties (\ref{property1}),(\ref{property2}), the
first (second) integral of this expression can be calculated by
closing the integration contour in the lower (upper) $p$-complex
half plane and avoiding with a semicircle the singularity at
$p=0$. In both cases, the leading contribution is given by the
pole with the smallest modulus. When $\pi/6<\eta <\pi/2$, these
poles are at $p=\pm i/(4\eta)$ and therefore the total
contribution reads
\begin{equation}
N\pi-2N^{1-\frac {c \pi}{2\eta}}C^{-\frac {c \pi}{2\eta}}e^{-\frac
{x'\pi}{2\eta}}+2\left (1-\frac {2\eta}{\pi}\right)N^{1+\frac
{\pi}{2\eta}\left (c -\frac {d}{2}\right)}\tan \frac {\pi
^2}{4\eta}\, C^{\frac {\pi c}{2\eta}}D^{-\frac {\pi
d}{4\eta}}e^{\frac {x'\pi}{2\eta}} \, .
\end{equation}
The first term comes from the semicircle around $p=0$. For
technical reasons we need to restrict further the domain of $\eta$
to within $\frac {\pi}{6}<\eta<\frac {\pi}{4}$, so that $\tan
\frac {\pi ^2}{4\eta}>0$, and choose
\begin{equation}
c=\frac{2\eta}{\pi} \, , \quad d=\frac{8\eta}{\pi} \, , \quad
C=\frac {4}{mR} \, , \quad D^2=\frac {16}{m^2R^2} \left (1-\frac
{2\eta}{\pi}\right)\tan \frac {\pi ^2}{4\eta}  \, , \label{scelta}
\end{equation}
where $m$ is a positive constant with the dimension of a mass and
$R$ is the lattice length. This choice entails
\begin{equation}
NF(x'+\Theta)=N\pi+mR\sinh \frac {\pi x'}{2\eta} +o(N^0) \, ,
\end{equation}
where $o(z)$ means ``order less than $z$". Of course, the
prescription (\ref{presc}) produces on the kernel function
$G(x'-y')$ (\ref{G2}) the result (\ref {G3}) as in the
trigonometric XXZ limit. Therefore, we obtain that the limiting
counting function $Z(x')$ (\ref {limzeta}) has to satisfy the
equation
\begin{equation}
Z(x')=mR\sinh \frac {\pi x'}{2\eta}+2 \int _{-\infty}^{+\infty
}dy' G_{sG}(x'-y') {\mbox {Im}}\ln \left [1+e^{iZ(y'+i0)}\right]
\, , \label {ddv4}
\end{equation}
where, since $N\in 4{\mathbb N}$, the constant $N\pi$ has been
reabsorbed in a redefinition of $Z(x')$ and eventually the
interval (\ref{rangexy}) has become infinite
\begin{equation}
-\infty <x',y'<+\infty  \, . \label{rengexyinfty}
\end{equation}
With the identification
\begin{equation}
\eta=\frac {\pi}{2}\left (1-\frac {b^2}{8\pi} \right ) \, ,
\label{etabiden}
\end{equation}
the Non-Linear Integral Equation (\ref{ddv4}) describes the vacuum
of the sine-Gordon field theory with coupling constant $b^2$ and
renormalised mass parameter $m$ (i.e. Lagrangian  ${\cal L}=\frac
{1}{2}(\partial \phi)^2+\frac {m_0^2}{b^2}\cos b \phi$) on a
(space-time) cylinder with spatial circumference $R$.

{\bf Remark 1} The Non-Linear Integral Equation (\ref{ddv4}) has
been obtained when $\pi/6<\eta<\pi/4$. However, it can be
considered without problems in the whole region $0<\eta<\pi/2$,
defining everywhere by analytical continuation the state which
minimises the energy.

{\bf Remark 2} We remark that in formula (\ref {force1}) one could
have made the choices $e^{-2i \frac {n \pi }{{\bf K}'}\left
(x+\frac {1}{2}{\bf K}' \right )} $ or $\frac {1}{2} e^{-2i \frac
{n \pi }{{\bf K}'}\left (x+\frac {1}{2}{\bf K}' \right )}+ \frac
{1}{2} e^{-2i \frac {n \pi }{{\bf K}'}\left (x-\frac {1}{2}{\bf
K}' \right )}$ in the exponential of the last term, in order to
express the factor $\cos n \pi$ in (\ref{force}). However, the
first choice together with the choice $x=x'-\Theta$,
$y=y'-\Theta$, leads to the equality
\begin{equation}
NF(x'-\Theta)=-N\pi+mR\sinh \frac {\pi x'}{2\eta} +o(N^0) \, .
\end{equation}
Instead, the second choice together with the shifts $x=x'\pm
\Theta$, $y=y'\pm \Theta$, gives
\begin{equation}
NF(x'\pm \Theta)=\pm \frac {1}{2} N\pi+mR\sinh \frac {\pi
x'}{2\eta} +o(N^0) \, .
\end{equation}
Since the constants $-N\pi$, $\pm \frac {1}{2}N\pi$ are inessential, we
eventually obtain again equation (\ref {ddv4}).

{\bf Remark 3} With the choices (\ref{property1}, \ref{property2},
\ref{scelta}), in the limit $N\rightarrow \infty$ the parameters
$J_x$, $J_y$, $J_z$ (\ref {Jpar}) behave as follows
\begin{eqnarray}
\frac {J_z}{J_x}&\rightarrow &\cos 2 \eta \, , \nonumber \\
\frac {J_x-J_y}{J_x \, 8 \sin ^2 2 \eta}&\simeq &\left
(\frac{MR}{4N}\right)^{\frac {8\eta}{\pi}} \rightarrow 0 \, ,
\label {Jpar2}
\end{eqnarray}
where we have introduced a rescaled mass parameter
\begin{equation}
M= \frac {m}{{\sqrt {\left |\left (1-\frac {2\eta}{\pi}\right)\tan \frac
{\pi ^2}{4\eta}\right | }}} \, .
\end{equation}
Clearly, the famous scaling limit to the sine-Gordon field theory
on the full spatial line \cite{LUT, JKM} is now gained from here
by sending $R\rightarrow \infty$.

{\bf Remark 4} The limit discussed in this section can be applied
to the eigenvalues (\ref{tra1}, \ref{tra2}) of the transfer
matrix. The calculations of the limits of the $Z$-independent
terms are less straightforward, but they are carried out following
a procedure analogous to that used on the forcing term
(remembering that due to the redefinition of $Z$ the forcing term
$F(x)$ has to be replaced by $F(x)-\pi$). Here we just give the
result:

$\bullet$ if $-2\eta < {\mbox {Im}}\alpha < 0$:
      \begin{eqnarray}
      \ln \Lambda ^+(\alpha)&=&-mR {\mbox {cotan}}\frac {\pi ^2}{4\eta}
\cosh \frac {\pi \alpha}{2\eta}+\nonumber \\
&& \\
      &+& \int _{-\infty}^{+\infty }dx \frac {1}{2\eta \, \sinh \frac
{\pi}{2 \eta}(\alpha -x) }{\mbox {Im}}\ln \left
[1+e^{iZ(x+i0)}\right] \,
      ;\nonumber
      \end{eqnarray}

$\bullet$ if $0< {\mbox {Im}}\alpha < \pi - 2\eta$:
\begin{eqnarray}
&& \ln \Lambda ^+(\alpha)=-mR \frac {1}{\sin \frac {\pi
^2}{4\eta}} \cosh \left [ \frac {\pi }{2\eta}\left (i\frac
{\pi}{2}-\alpha
\right)\right ]-\nonumber \\
&& \\
      &-& \int _{-\infty}^{+\infty }dx \int _{-\infty}^{+\infty } \frac
{dp}{2\pi}\frac {i\sinh p\eta}{\sinh p \left (\frac
{\pi}{2}-\eta\right )\cosh p\eta}e^{ip\left (\alpha+i\eta-i\frac
{\pi}{2}-x\right)}{\mbox {Im}}\ln \left [1+e^{iZ(x+i0)}\right]
      , \nonumber
      \end{eqnarray}
which hold when $\pi/4<\eta<\pi/2$. Thanks to a sign change of
$\alpha$, these unveil the coincidence with the vacuum eigenvalue
of the transfer matrix in the sine-Gordon theory,
$\Lambda_{sG}(\alpha)$, as computed in the second of \cite{FR1}
\begin{equation}
\ln \Lambda ^+(\alpha)=\ln \Lambda ^+_{sG}(-\alpha) \, .
\end{equation}

\section{Scattering theory}
\setcounter{equation}{0} As the spin chain becomes infinitely
long, $N\rightarrow \infty$ and fixed $\Delta$, the infra-red (IR)
Hamiltonian eigenstates become on mass-shell states. Besides the
statistical mechanics interpretation as thermodynamic limit
configurations, these may also be thought of as describing the
asymptotic particles in the quantum field theory perspective. And
in the same view the excited state version of the nonlinear
(elliptic) integral equation (\ref{ddv3}), being merely a
quantisation rule, gives exactly the scattering amplitudes of the
asymptotic states provided that the convolution integral is
neglected (when $N\rightarrow \infty$). Moreover, integrability
reveals itself useful in the factorisation of the general
scattering S-matrix into two particle ones, which eventually
encode all the on mass-shell information. Actually, this kind of
approach was bashfully commenced in \cite{FMQR} and seriously
pursued in \cite{BOL} for what concerns the sine-Gordon theory.

We shall distinguish two cases: 1) the repulsive regime $0<\eta
<\frac {\bf K}{2}$ (from Section 5 this means indeed
$4\pi<b^2<8\pi$ in the sine-Gordon limit); 2) the attractive case
$\frac {\bf K}{2}<\eta<{\bf K}$ (which covers the remainder
$0<b^2<4\pi$). In the first case we only have soliton and
antisoliton excitations; in the second case we also have
soliton-antisoliton bound states, the so-called breathers: the
difference with sine-Gordon is that everything become {\it
deformed} or {\it elliptic}. In general, the key ingredient shall
be the primitive function of the integral kernel (\ref{G2}),
suitably normalised in the rapidity variable $\tilde \theta$ as
\begin{equation}
\chi (\tilde \theta)=\int _0^{\frac {\eta}{\bf K}\tilde \theta} dx
2 \pi G(x)=i\sum _{\stackrel {n=-\infty}{n \not= 0}}^{+\infty}
\frac {\sinh \frac {2n({\bf K}-2\eta)\pi}{{\bf K}'}}{2\sinh \frac
{2n({\bf K}-\eta)\pi}{{\bf K}'} \cosh \frac {2n\eta \pi}{{\bf
K}'}}  \frac {e^{-2i \frac {n \pi \eta \tilde \theta}{{\bf K}{\bf
K}'}}}{n} \, , \label{chi}
\end{equation}
which holds within the domain $|{\mbox {Im}}\tilde \theta |<2{\bf
K}$. In the previous equality, we preferred to use the {\it
renormalised} rapidity $\tilde \theta$, which is connected to the
{\it bare} rapidity $x$ used in (\ref {ddv3}) by the relation
\begin{equation}
\tilde \theta=\frac {\bf K}{\eta}x \, . \label{renbare}
\end{equation}
In fact, the excited state version of the NLIE (\ref{ddv3}) can be
now written in a more compact form in terms of $\tilde \theta$ by
following the pattern of \cite{FMQR}:
\begin{eqnarray}
Z_N(\tilde \theta)&=&N F(\tilde \theta) +\sum _{k=1}^{N_h}\chi
(\tilde \theta -h_k) -\sum _{k=1}^{N_c}\chi (\tilde \theta -c_k)
-\sum _{k=1}^{N_w}\chi _{II}(\tilde \theta -w_k)  -\sum
_{k=1}^{N_{sc}}\chi _{II}(\tilde
\theta -s_k)+ \nonumber \\
&+&2 \int _{-\frac {{\bf K}'}{2}}^{\frac {{\bf K}'}{2}}d\tilde
\eta  \bar G(\tilde \theta-\tilde \eta) {\mbox {Im}}\ln \left
[1+e^{iZ_N(\tilde \eta+i0)}\right] \, , \label{ddvex}
\end{eqnarray}
for an excitation with $N_h$ (real) holes, $N_c$ close (complex)
pairs, $N_w$ wide (complex) pairs  and $N_{sc}$ (complex)
self-conjugated roots.\footnote{Generally, the roots come in
complex-conjugated pairs because of the real analyticity of the
counting function $Z(x)$. A pair of complex conjugate roots is
said to be close if the absolute value of their imaginary parts is
$<2 {\bf K}$, -- i.e. the domain of the function $\chi(\tilde
\theta)$ (\ref{chi}) --, wide if it is $>2 {\bf K}$. Nevertheless,
one root may be {\it single} if and only if it is either real or
self-conjugated. In other words, they behave exactly like
sine-Gordon roots \cite{FMQR}, although the fixed imaginary part
of one self-conjugated root will be $-{{\bf K}^2}/\eta$.} The
reason for the appearance of the function $\chi _{II}$, the second
determination of $\chi$, is an important technical detail which
will be discussed in what follows. We have also introduced a
different normalisation for the kernel: $\bar G(\tilde \theta)=
\frac {\eta}{\bf K} G \left (\frac {\eta}{\bf K}\tilde \theta
\right )$. For characterising the state, additional algebraic
equations on $Z_N(\tilde \theta)$ would also be necessary
\cite{FMQR}; but these do not affect directly our scattering
treatment and then are left out in this context.

\subsection {Repulsive regime: $0<\eta <\frac {\bf K}{2}$}

In this regime of the XYZ model, the two particle asymptotic
states are spanned by four independent vectors, which describe
respectively the soliton-soliton, the antisoliton-soliton, the
soliton-antisoliton and the antisoliton-antisoliton excitations.
Although the two-particle S-matrix is, as said before, the
building block, we may consider, in general and without extra
difficulties, the multi-particle states as those forming a
representation space of the Zamolodchikov-Faddeev algebra. In
fact, this is a non-commutative algebra with generators
$A(\theta)$ and $\bar A(\theta)$ satisfying
\begin{eqnarray}
A(\theta _1)A(\theta _2)&=&S(\theta _1-\theta _2) A(\theta
_2)A(\theta _1)+S_a(\theta _1-\theta _2) \bar A(\theta
_2)\bar A(\theta _1) \, \\
A(\theta _1)\bar A(\theta _2)&=&S_t(\theta _1-\theta _2)\bar
A(\theta _2)A(\theta _1)+S_r(\theta _1-\theta _2) A(\theta
_2)\bar A(\theta _1) \, \\
\bar A(\theta _1)A(\theta _2)&=&S_t(\theta _1-\theta _2) A(\theta
_2)\bar A(\theta _1)+S_r(\theta _1-\theta _2) \bar A(\theta
_2) A(\theta _1) \, \\
\bar A(\theta _1)\bar A(\theta _2)&=&S(\theta _1-\theta _2) \bar
A(\theta _2)\bar A(\theta _1)+S_a(\theta _1-\theta _2) A(\theta
_2)A(\theta _1) \, .\label{FZalg}
\end{eqnarray}
Physically, the generator $A(\theta)$ ($\bar A(\theta)$) describes
a soliton (antisoliton) excitation on the vacuum with rapidity
$\theta$. Of course, this heuristic consideration leads
immediately to the mentioned particle representations of this
algebra: asymptotic in (out) states are created on the vacuum by
products in which the generators are arranged with decreasing
(increasing) rapidities. As a particular case, the scattering in
the two-particle sector may be algebraically described by the four
state basis mentioned at the beginning of this Subsection and the
functions $S(\theta_{12})$, $S_a(\theta_{12})$,
$S_r(\theta_{12})$, $S_t(\theta_{12})$, where
$\theta_{12}=\theta_1-\theta_2$, selected as the only
non-vanishing elements of the two-particle S-matrix:
\begin{eqnarray}
_{out}<\bar A(\theta _1)\bar A(\theta _2) | \bar A(\theta _1)\bar
A(\theta _2) >_{in}=_{out}<A(\theta _1)A(\theta _2) | A(\theta
_1)A(\theta _2) >_{in} = S(\theta_{12}) \, , \\
_{out}<\bar A(\theta _1)A(\theta _2) | \bar A(\theta _1) A(\theta
_2)>_{in}=_{out}<A(\theta _1)\bar A(\theta _2) | A(\theta _1)\bar
A(\theta
_2) >_{in} = S_t(\theta _{12}) \, , \\
_{out}<\bar A(\theta _1)A(\theta _2) | A(\theta _1) \bar A(\theta
_2)>_{in}=_{out}<A(\theta _1)\bar A(\theta _2) | \bar A(\theta
_1)A(\theta
_2) >_{in} = S_r(\theta_{12}) \, , \\
_{out}<\bar A(\theta _1)\bar A(\theta _2) | A(\theta _1)A(\theta
_2)>_{in}=_{out}<A(\theta _1)A(\theta _2) | \bar A(\theta _1)\bar
A(\theta _2)>_{in} = S_a(\theta_{12}) \, . \label{AA}
\end{eqnarray}
     From these considerations, it follows that each of the following
four states
\begin{equation}
\quad \frac {1}{\sqrt {2}}(|A(\theta _1)\bar A(\theta _2)>_{in}\pm
|\bar A(\theta _1)A(\theta _2)>_{in}) \, , \frac{1}{\sqrt
{2}}(|A(\theta _1)A(\theta _2)>_{in}\pm |\bar A(\theta _1) \bar
A(\theta _2)>_{in})   \label{Seig}
\end{equation}
is preserved by the scattering, i.e. it is an eigenvector of the
two particle S-matrix. The respective eigenvalues are the four
amplitudes of the scattering processes
\begin{eqnarray}
\frac{1}{2}\cdot_{out}<A(\theta _1)\bar A(\theta _2)\pm \bar
A(\theta _1) A(\theta _2) | A(\theta _1)\bar A(\theta _2)\pm \bar
A(\theta _1) A(\theta _2)>_{in} = S_t(\theta _{12})\pm S_r(\theta
_{12}) \, , \nonumber \\
\label{Seigenvalues} \\
\frac{1}{2}\cdot_{out}<A(\theta _1)A(\theta _2)\pm \bar A(\theta
_1) \bar A(\theta _2) | A(\theta _1)A(\theta _2) \pm \bar A(\theta
_1) \bar A(\theta _2)>_{in} = S(\theta _{12})\pm S_a(\theta _{12})
\, . \nonumber
\end{eqnarray}
As a matter of fact, we have written down the eigenvalues of the
two-particle S-matrix because we will read off exactly them from
the IR limit ($N\rightarrow \infty$) of the excited state NLIE
(\ref{ddvex}).

Let us start by considering the first eigenvector
\begin{equation}
\frac {1}{\sqrt {2}}(|A(\theta _1)\bar A(\theta _2)>_{in} +|\bar
A(\theta _1) A(\theta _2)>_{in}) \, , \label{antsol}
\end{equation}
which describes the symmetric combination of a soliton and an
antisoliton, and by computing its eigenvalue
\begin{equation}
S_+(\theta)= S_t(\theta )+S_r(\theta ) \, , \label{S-}
\end{equation}
with the shorter definition $\theta=\theta_{12}$. As Bethe Ansatz
state it is built up by exciting  two holes and a (close) complex
pair of roots upon the (Fermi-Dirac) real root sea. Now, this
configuration entails in a standard manner \cite{FMQR} an excited
state nonlinear integral equation of the form (\ref{ddvex}) with a
contribution of four $\chi(\tilde \theta- \tilde \theta_i)$
($\tilde \theta _i=h_1,h_2,c,\bar{c}$ denotes the two holes or the two
complex
roots respectively). However, as far as the scattering factor is
concerned, these four become three once $\tilde \theta$ equals any
of the hole rapidities. Decisive simplifications occur as $N$
grows: the imaginary parts of the complex roots approach $\pm {\bf
K}$ respectively, since the expression (\ref{chi}) has exactly
poles at $|{\mbox {Im}}\tilde \theta |=2{\bf K}$, and the real
part tends to the hole middle point (as anticipated the
convolution integral becomes negligible). Thanks to all these
facts, we can identify with $\theta$ the difference between the
hole positions and write down the scattering amplitude as the
exponential of the new contribution
\begin{equation}
S_+(\theta)={\mbox {exp}}\left [ -i \chi \left (\frac
{\theta}{2}-i{\bf K} \right)  -i \chi \left (\frac
{\theta}{2}+i{\bf K} \right)\right]  {\mbox {exp}}\left [ i \chi
(\theta)\right]  \, .
\end{equation}
Upon exploiting (\ref{chi}), this scattering amplitude may be
manipulated into
\begin{equation}
S_+(\theta)={\mbox {exp}}\left [ -\sum _{n=1}^{\infty}\frac {2i
\sinh \frac {2n({\bf K}-2\eta)\pi}{{\bf K}'} \sin \frac {n \pi
\eta \theta}{{\bf K}{\bf K}'}}{n \sinh \frac {2n({\bf
K}-\eta)\pi}{{\bf K}'} }\right] {\mbox {exp}}\left [ \sum
_{n=1}^{\infty}\frac {i \sinh \frac {2n({\bf K}-2\eta)\pi}{{\bf
K}'} \sin \frac {2n \pi \eta \theta}{{\bf K}{\bf K}'}}{n \sinh
\frac {2n({\bf K}-\eta)\pi}{{\bf K}'} \cosh \frac {2n\eta
\pi}{{\bf K}'}} \right ] \, , \label {S-1}
\end{equation}
or expressed in a more compact manner thanks to the function
$\theta _{11}$ (\ref {thetafunc}) with nome ${\mbox {exp}}\left (
-2\pi\frac {{\bf K}-\eta}{{\bf K}'}\right )$:
\begin{equation}
S_+(\theta)=-e^{\frac {i\pi \eta \theta}{{\bf K}{{\bf K}'}} }\frac
{\theta _{11}\left (-\frac {\eta \theta}{2{\bf K}{\bf K}'}-i\frac
{\eta}{{\bf K}'} ; 2i \frac {{\bf K}-\eta}{{\bf K}'} \right )}
{\theta _{11}\left (-\frac {\eta \theta}{2{\bf K}{\bf K}'}+i\frac
{\eta}{{\bf K}'} ; 2i \frac {{\bf K}-\eta}{{\bf K}'} \right )}
{\mbox {exp}}\left [ \sum _{n=1}^{\infty}\frac {i \sinh \frac
{2n({\bf K}-2\eta)\pi}{{\bf K}'} \sin \frac {2n \pi \eta
\theta}{{\bf K}{\bf K}'}}{n \sinh \frac {2n({\bf K}-\eta)\pi}{{\bf
K}'} \cosh \frac {2n\eta \pi}{{\bf K}'}} \right ] \, . \label
{S-2}
\end{equation}
Let us pass on to the antisymmetric state in the same
soliton-antisoliton sector
\begin{equation}
\frac {1}{\sqrt {2}}(|A(\theta _1)\bar A(\theta _2)>_{in} - |\bar
A(\theta _1) A(\theta _2)>_{in}) \, , \label{symsol}
\end{equation}
and find out its eigenvalue
\begin{equation}
S_-(\theta)= S_t(\theta )- S_r(\theta ) \, . \label{S+}
\end{equation}
It may be analogously described by a configuration with two holes
and a wide pair of complex roots and then implies a four $\chi$
contribution to the r.h.s. of the excited equation (\ref{ddvex}).
While $N$ is growing the two roots tend to be a single
self-conjugate wide root with imaginary part $-{{\bf K}^2}/\eta$
and real part again fixed by the hole middle point. In fact, this
root lies outside the strip $|{\mbox {Im}}\tilde \theta |<2 {\bf
K}$ (here $0<\eta<{\bf K}/2$) and therefore we need to extend
$\chi(\tilde \theta)$ (\ref{chi}) into the so-called second
determination, which is given (up to a constant) by
\begin{equation}
\chi_{II}(\tilde \theta)=\chi(\tilde \theta)+\chi(\tilde
\theta-{\mbox {sgn}}({\mbox {Im}}\tilde \theta)2i{\bf {K}}) \, .
\end{equation}
Therefore, (\ref{chi}) yields explicitly
\begin{equation}
\chi _{II}(\tilde \theta)=\frac {1}{i}\log \left [ \frac {\theta
_{01}\left (\frac {\eta \tilde \theta}{{\bf K}{\bf K}'}-{\mbox
{sgn}}({\mbox {Im}} \tilde \theta)i\frac {\eta}{{\bf K}'}+i\frac
{{\bf K}-2\eta}{{\bf K}'} ; 2i \frac {{\bf K}-\eta}{{\bf K}'}
\right )} {\theta _{01}\left (\frac {\eta \tilde \theta}{{\bf
K}{\bf K}'}-{\mbox {sgn}}({\mbox {Im}} \tilde \theta)i\frac
{\eta}{{\bf K}'}-i\frac {{\bf K}-2\eta}{{\bf K}'} ; 2i \frac {{\bf
K}-\eta}{{\bf K}'} \right )} {\mbox {exp}}\left (-2\pi \frac {{\bf
K}-2\eta}{\bf K'}{\mbox {sgn}}{\mbox {Im}}\tilde \theta \right)
\right ] \, , \label {chi2}
\end{equation}
through the entire domain of second determination
\begin{equation}
2{\bf K}<|{\mbox {Im}}\tilde \theta |<\frac {2{\bf
K}^2}{\eta}-2{\bf K} \label{chi2regio},
\end{equation}
which is the range of variation of the imaginary part too: $2{\bf
K}<{\bf K}^2/\eta<2{\bf K}^2/\eta-2{\bf K}$ (as $0<\eta <{\bf
K}/2$). Therefore, with the same definition of $\theta$ the
exponential of the new contribution reads
\begin{equation}
S_-(\theta)={\mbox {exp}}\left [ -i \chi _{II}\left (\frac
{\theta}{2}-i\frac {{\bf K}^2}{\eta} \right) \right]  {\mbox
{exp}}\left [ i \chi (\theta)\right]  \, ,
\end{equation}
or equivalently if we make use of (\ref {chi2})
\begin{equation}
S_-(\theta)= - e^{\frac {i\pi \eta \theta}{{\bf K}{{\bf K}'}} }
\frac {\theta _{01}\left (-\frac {\eta \theta}{2{\bf K}{\bf
K}'}-i\frac {\eta}{{\bf K}'} ; 2i \frac {{\bf K}-\eta}{{\bf K}'}
\right )} {\theta _{01}\left (-\frac {\eta \theta}{2{\bf K}{\bf
K}'}+i\frac {\eta}{{\bf K}'} ; 2i \frac {{\bf K}-\eta}{{\bf K}'}
\right )} {\mbox {exp}}\left [ \sum _{n=1}^{\infty}\frac {i \sinh
\frac {2n({\bf K}-2\eta)\pi}{{\bf K}'} \sin \frac {2n \pi \eta
\theta}{{\bf K}{\bf K}'}}{n \sinh \frac {2n({\bf K}-\eta)\pi}{{\bf
K}'} \cosh \frac {2n\eta \pi}{{\bf K}'}} \right ] \, . \label
{S+1}
\end{equation}
Now, we want to highlight that we have chosen the undetermined
constant factor in the definition of $\chi(\tilde \theta)$ and
$\chi _{II}(\tilde \theta)$ in such a way that $S_+(\theta =0)=1$,
$S_-(\theta=0)=-1$: this is indeed what happens for the limiting
amplitudes in sine-Gordon.

At this point, we are left with the calculation of the remaining
two eigenvalues, $S_1(\theta)=S(\theta)+S_a(\theta)$,
$S_2(\theta)=S(\theta)-S_a(\theta)$, corresponding respectively to
the two second eigenvectors of (\ref{Seig}):
\begin{equation}
\frac{1}{\sqrt {2}}(|A(\theta _1)A(\theta _2)>_{in}\pm |\bar
A(\theta _1) \bar A(\theta _2)>_{in}) \, .
\end{equation}
But now the similarity with the sine-Gordon configurations clearly
fails. In fact, there \cite{FMQR} the antisymmetric combination is
simply realised by a Bethe state with two holes and hence would
breed a two $\chi(\tilde\theta- \tilde\theta_j)$ ($j=1,2$)
contribution in the r.h.s. of (\ref{ddvex}). And in sine-Gordon
theory it gives the eigenvalue of the symmetric combination as
well. In other words, both eigenvalues coincide with the
scattering amplitude soliton-soliton into soliton-soliton there,
being the soliton-soliton into antisoliton-antisoliton event
prevented by topological charge conservation ($S_a=0$). Yet, in
the XYZ chain the situation is richer and this degeneracy removed:
in physical language the channel soliton-soliton into
antisoliton-antisoliton is here allowed and the topological charge
not conserved but modulo $4$. As a consequence, we need to go
along a different route which shall be connected with the
existence of a novelty in this scenario: the real periodicity.
And indeed, in the elliptic case, we may also consider
configurations with two holes and a (close or wide) complex pair
in which the position of the real part of the complex pair
undergoes a shift by $\pm \frac {1}{2} {\bf K}'$ (in the $x$
coordinate) from the middle point of the holes. In terms of
$\theta=\frac {\bf K}{\eta}x$, the scattering factors deriving
from these new configurations are given by (\ref{S-2}) and (\ref
{S+1}) (respectively for symmetric and antisymmetric state) with
$\theta$ replaced by $\theta \pm \frac {{\bf K}{\bf K}'}{\eta}$
(the hole contribution ${\mbox {exp}}[i\chi (\theta)]$, which
should not change, is indeed invariant under those shifts of
$\theta$). These simplify further after using $\theta _{11}(z\pm
1/2;\tau )= \mp \theta _{10}(z;\tau )$ and $\theta
_{01}(z\pm1/2;\tau )= \theta _{00}(z;\tau )$ respectively:
\begin{eqnarray}
S_1(\theta)&=&e^{\frac {i\pi \eta \theta}{{\bf K}{{\bf K}'}}
}\frac {\theta _{10}\left (-\frac {\eta \theta}{2{\bf K}{\bf
K}'}-i\frac {\eta}{{\bf K}'} ; 2i \frac {{\bf K}-\eta}{{\bf K}'}
\right )} {\theta _{10}\left (-\frac {\eta \theta}{2{\bf K}{\bf
K}'}+i\frac {\eta}{{\bf K}'} ; 2i \frac {{\bf K}-\eta}{{\bf K}'}
\right )} {\mbox {exp}}\left [ \sum _{n=1}^{\infty}\frac {i \sinh
\frac {2n({\bf K}-2\eta)\pi}{{\bf K}'} \sin \frac {2n \pi \eta
\theta}{{\bf K}{\bf K}'}}{n \sinh \frac {2n({\bf K}-\eta)\pi}{{\bf
K}'} \cosh \frac {2n\eta \pi}{{\bf K}'}} \right ] \, , \nonumber
\\
\label {S+-}\\
S_2(\theta)&=& e^{\frac {i\pi \eta \theta}{{\bf K}{{\bf K}'}} }
\frac {\theta _{00}\left (-\frac {\eta \theta}{2{\bf K}{\bf
K}'}-i\frac {\eta}{{\bf K}'} ; 2i \frac {{\bf K}-\eta}{{\bf K}'}
\right )} {\theta _{00}\left (-\frac {\eta \theta}{2{\bf K}{\bf
K}'}+i\frac {\eta}{{\bf K}'} ; 2i \frac {{\bf K}-\eta}{{\bf K}'}
\right )} {\mbox {exp}}\left [ \sum _{n=1}^{\infty}\frac {i \sinh
\frac {2n({\bf K}-2\eta)\pi}{{\bf K}'} \sin \frac {2n \pi \eta
\theta}{{\bf K}{\bf K}'}}{n \sinh \frac {2n({\bf K}-\eta)\pi}{{\bf
K}'} \cosh \frac {2n\eta \pi}{{\bf K}'}} \right ] \, . \nonumber
\end{eqnarray}

\medskip

Up to now, we found the scattering amplitudes in the basis vectors
describing two particle states (of solitons and antisolitons).
However, in order to find a link with the famous Baxter's
eight-vertex R-matrix (\cite{BAX} and \cite{BAXbook}), we need to
describe the same scattering process in an equivalent way by using
a different basis. In this manner, we will also deduce the
Zamolodchikov's S-matrix of \cite {ZAMe}. In fact, we may
introduce the real doublet of particles, $A_1$ and $A_2$, as
\begin{equation}
A(\theta)=\frac {1}{\sqrt {2}}[A_1(\theta)+iA_2(\theta)] \, ,
\quad \bar A(\theta)=\frac {1}{\sqrt
{2}}[A_1(\theta)-iA_2(\theta)] \, .
\end{equation}
In terms of these new generators the Zamolodchikov-Faddeev algebra
(\ref{FZalg}) looks as follows
\begin{eqnarray}
A_1(\theta _1)A_1(\theta _2)&=&\sigma (\theta _1-\theta _2)
A_1(\theta _2)A_1(\theta _1)+\sigma_a(\theta _1-\theta _2)
A_2(\theta
_2) A_2(\theta _1) \, \\
A_1(\theta _1)A_2(\theta _2)&=&\sigma_t(\theta _1-\theta _2)
A_2(\theta _2)A_1(\theta _1)+\sigma_r(\theta _1-\theta _2)
A_1(\theta
_2)A_2(\theta _1) \, \\
A_2(\theta _1)A_2(\theta _2)&=&\sigma (\theta _1-\theta _2)
A_2(\theta _2)A_2(\theta _1)+\sigma_a(\theta _1-\theta _2)
A_1(\theta
_2) A_1(\theta _1) \, \\
A_2(\theta _1)A_1(\theta _2)&=&\sigma_t(\theta _1-\theta _2)
A_1(\theta _2)A_2(\theta _1)+\sigma_r(\theta _1-\theta _2)
A_2(\theta _2)A_1(\theta _1) \, , \label{FZalg2}
\end{eqnarray}
where the $\sigma $-amplitudes are related to the S-amplitudes by
means of the relations
\begin{eqnarray}
2 \sigma = S_t+S_r+S+S_a \, , \nonumber \\
2 \sigma _a= S_t+S_r-S-S_a \, , \nonumber \\
2 \sigma _t= S+S_t-S_a-S_r \, , \nonumber \\
2 \sigma _r= S+S_r-S_a-S_t \, . \label {sigma-S}
\end{eqnarray}
Namely, the eigenvalues are given by
\begin{eqnarray}
S_+=S_t+S_r=\sigma +\sigma _a \, , \nonumber \\
S_-=S_t-S_r=\sigma _t -\sigma _r \, , \nonumber \\
S_1=S+S_a=\sigma -\sigma _a \, , \nonumber \\
S_2=S-S_a=\sigma _t +\sigma _r \, .  \label {eigSsigma}
\end{eqnarray}
 From the expressions for $S_+$, $S_-$, $S_1$, $S_2$ (\ref {S-2},
\ref {S+1}, \ref {S+-}) and after some lenghty calculation, we
obtain the $\sigma$-amplitudes,
\begin{eqnarray}
\sigma (\theta)&=&\frac {(x;P,Q^4)^2
(Q^2x^{-1};P,Q^4)^2}{(x^{-1};P,Q^4)^2 (Q^2x;P,Q^4)^2} \cdot
\label{ctheta} \\
&\cdot & \frac
{(x^{-1};P)(x^{-1};Q^4)(Q^2x;Q^4)(x^{-1}P;P^2)(xP;P^2)(Q^2;P^2)
(P^2Q^{-2};P^2)}
{(x;P)(x;Q^4)(Q^2x^{-1};Q^4)(x^{-1}Q^2;P^2)(P^2Q^{-2}x;P^2)(P;P^2)^2}
\, , \nonumber
\end{eqnarray}
where, with a {\it new definition} of $x$ which holds hereafter
\footnote{Because of the heavy technicality, we preferred to keep
the symbol mainly used in the current literature at the cost of
this notation abuse.},
\begin{equation}
x={\mbox {exp}}\left ( \frac {2i\eta \pi \theta}{\bf {K}\bf {K}'}
\right ) \, , \quad Q={\mbox {exp}}\left ( -\frac {2\eta \pi}{\bf
{K}'} \right ) \, , \quad P={\mbox {exp}}\left ( 4\frac {\eta -
{\bf {K}}}{\bf {K}'}\pi \right ) \, , \label{QP}
\end{equation}
and also the remaining ones
\begin{eqnarray}
\frac {\sigma _a(\theta)}{\sigma (\theta)}=\frac {{\mbox
{sn}}\left (\frac {2\eta \theta K}{\bf {K}\bf {K}'}; 4i \frac
{{\bf K}-\eta}{{\bf K}'}\right )} {{\mbox {sn}}\left (\frac
{4i\eta K}{\bf {K}'}; 4i \frac {{\bf K}-\eta}{{\bf K}'}\right )}\,
, \quad \frac{\sigma _r(\theta)}{\sigma (\theta)}=-\frac {{\mbox
{sn}}\left (\frac {2\eta \theta K}{\bf {K}\bf {K}'}-\frac {4i\eta
K}{\bf {K}'}; 4i \frac {{\bf K}-\eta}{{\bf K}'}\right)} {{\mbox
{sn}}\left ( \frac {4i\eta K}{\bf {K}'}; 4i \frac {{\bf
K}-\eta}{{\bf K}'}\right ) }\,
,\nonumber \\
\frac {\sigma _t(\theta)}{\sigma (\theta)}=k\,  {{\mbox {sn}}\left
(\frac {2\eta \theta K}{\bf {K}\bf {K}'}-\frac {4i\eta K}{\bf
{K}'}; 4i \frac {{\bf K}-\eta}{{\bf K}'}\right)} {{\mbox {sn}}
\left (\frac {2\eta \theta K}{\bf {K}\bf {K}'}; 4i \frac {{\bf
K}-\eta}{{\bf K}'}\right )} \, . \label{abcd2}
\end{eqnarray}
Of course, the Jacobian elliptic function sn has nome $P$ now and
also $K$ and $k$ are respectively the first kind complete elliptic
integral and the modulus corresponding to the same nome $P$.

We are now ready to identify our new $\sigma$-amplitudes with the
entries of the eight-vertex R-matrix. Moreover, Baxter's R-matrix
was used more recently in order to define the elliptic algebra
${\cal A}_{q_e,p_e}(\widehat {sl}(2)_c)$ \cite {FIJKMY}. More
precisely, either the Baxter's R-matrix $R_B(x_e;q_e,p_e)$
(defined by (22) in \cite{FIJKMY}) and the R-matrix with scaled
nome $R_B^*(x_e;q_e,p_e)=R_B(x_e;q_e,p_e^*=p_eq_e^{-2c})$ are
involved in the definition of this elliptic algebra. Therefore the
relevance of this identification, which goes through the simple
definition
\begin{equation}
S_{rep}(\theta )=\left ( \begin{array}{cccc}
\sigma _r(\theta ) & 0 & 0 & \sigma _t(\theta ) \\
0 & \sigma _a(\theta ) & \sigma (\theta ) & 0 \\
0 & \sigma (\theta ) & \sigma _a(\theta ) & 0 \\
\sigma _t(\theta ) & 0 & 0 & \sigma _r(\theta ) \end{array} \right
) \, .\label{Sbax}
\end{equation}
As usual we shall parametrise the elliptic affine parameter $x_e$
by an exponential mapping of the physical rapidity
\begin{equation}
x_e^{-2}=x={\mbox {exp}}\left (\frac {2i\pi \eta \theta}{\bf {K}
\bf {K}'} \right ) \, . \label{affphy}
\end{equation}
Furthermore, we need to relate the deformation parameters, $q_e=Q$
and $p_eq_e^{-2}=P$, to finalise our link
\begin{equation}
R_B^*(x_e;q_e,p_e)|_{c=1}=S_{rep}(\theta) \, . \label{smir}
\end{equation}
Regarding the previous relation it is worth saying that the
authors of \cite{FIJKMY} highlighted the impossibility to find the
S-matrix of the XYZ chain in the literature, but reported a
conjecture due to F. Smirnov according to which it should be given
by $-R^*_B(x_e; q_e, p_e)|_{c=1}$. In this respect, we are now in
the position to unveil the mapping between the algebra parameters
$q_e$, $p_e$ and the physical variables of the XYZ chain,
\begin{equation}
q_e={\mbox {exp}}\left ( -\frac {2\eta \pi}{\bf {K}'} \right ) \,
, \quad p_e={\mbox {exp}}\left ( -\frac {4\pi {\bf {K}}}{\bf {K}'}
\right ) \, , \label{qepe}
\end{equation}
along with the previous relation (\ref{affphy}) concerning the
rapidity. As a check, we can verify that the range of parameters
considered in \cite{FIJKMY},
\begin{equation}
0<p_e < q_e^4 \, ,
\end{equation}
corresponds indeed to the repulsive regime.

As a second comparison, we want to show up the way of relating our
results on XYZ scattering factors to the $Z_4$-symmetric S-matrix
obtained by A.B. Zamolodchikov \cite{ZAMe}. The latter was derived
as an elliptic solution of the factorization (Yang-Baxter),
unitarity and analyticity conditions depending on the rapidity
$\theta _z$ and the parameters $\gamma $ and $\gamma '$. Once we
identify these with the XYZ chain rapidity and parameters,
respectively, in this manner
\begin{eqnarray}
{\mbox {exp}}\left ( \frac {2i\pi \eta \theta}{\bf {K} \bf {K}'}
\right )={\mbox {exp}}\left ( \frac {4i\pi \theta _z}{\gamma '}
\right ) \, &,& \quad {\mbox {exp}}\left ( -\frac {4\eta \pi}{\bf
{K}'} \right )={\mbox
{exp}}\left ( -\frac {4\pi ^2}{\gamma '} \right ) \, , \nonumber \\
{\mbox {exp}}\left ( 4\frac {\eta - {\bf {K}}}{\bf {K}'}\pi \right
) &=& {\mbox {exp}}\left ( -4\pi \frac {\gamma }{\gamma '}\right )
\, , \label{zamxyz}
\end{eqnarray}
we obtain that our $\sigma $, $\sigma _a$, $\sigma _t$ and
$\sigma_r$ coincide with the homonymous quantities in \cite{ZAMe}.

The last due comparison is with a Takebe's work \cite{TAK}, which
apparently do not contain mention to Zamolodchikov's S-matrix.
This paper contains several subtleties which make the verification
of its rightness quite difficult (and probably did not contribute
to its diffusion). Nevertheless, the method is remarkably
interesting and describes XYZ chain states directly in the $N
\rightarrow \infty$ limit by converting Bethe equations into a
linear equation for the ``density" of roots. Scattering factors
for excited states are then read off from the $o(1)$
next-to-leading contribution to the phase shift. The latter arises
after a complete circulation of the spatial (periodic) direction.
If Takebe's parameters are expressed in terms of ours as
\begin{equation}
\lambda _{T} =\frac {i\alpha}{2{\bf K}} \, , \quad \tau_{T}=\frac
{i{\bf K}'}{2{\bf K}} \, , \quad \eta _{T}=\frac {\eta}{2{\bf K}}
\, , \label {identi}
\end{equation}
it is rather an easy matter to reformulate our scattering factors
in terms of those in (2.3.54) of \cite{TAK}:
\begin{equation}
S_+(\theta)=(2.3.54)/1 \, , \quad S_-(\theta)=(2.3.54)/3 \, ,
\quad -S_1(\theta)=(2.3.54)/2 \, , \quad -S_2(\theta)=(2.3.54)/4
\, , \label{Takus}
\end{equation}
where (2.3.54) exploits in \cite{TAK} the new variables $x
_{T}=-\frac {\lambda _{T}}{\tau _{T}}$ and $t _{T}=\frac
{i}{\tau_{T}}$.

\medskip

Finally, in order to have a check on our results, it is important
to analyse what happens to the XYZ repulsive S-matrix in the
scaling sine-Gordon limit, i.e. ${\bf K'} \rightarrow \infty$,
which implies ${\bf K}\rightarrow \pi/2$ (cfr. Section 5). In this
context, thanks to the useful trigonometric limit
\begin{equation}
\lim _{{\bf K'} \rightarrow \infty} {\mbox {sn}}\left (\frac
{2\eta \theta K}{\bf {K}\bf {K}'}; 4i \frac {{\bf K}-\eta}{{\bf
K}'}\right )= \frac {\sinh \frac {\eta \theta}{\pi -2\eta}}{\cosh
\frac {\eta \theta}{\pi -2\eta}} \, , \label{snlim}
\end{equation}
we may easily obtain the sine-Gordon values
\begin{eqnarray}
S_+(\theta)&=&S_t(\theta)+S_r(\theta)\rightarrow - \frac {\sinh
\frac {\eta (\theta +i\pi)}{\pi -2\eta}}{\sinh \frac {\eta (\theta
-i\pi)}{\pi -2\eta}}e^{i\chi _{sG}(\theta)} \, ,
\nonumber \\
S_-(\theta)&=&S_t(\theta)-S_r(\theta)\rightarrow  -\frac {\cosh
\frac {\eta (\theta +i\pi)}{\pi -2\eta}}{\cosh \frac {\eta (\theta
-i\pi)}{\pi -2\eta}}e^{i\chi _{sG}(\theta)} \, ,
\label{abcdlim} \\
S_1(\theta)&=&S(\theta)+S_a(\theta)\rightarrow  e^{i\chi
_{sG}(\theta)} \, , \nonumber \\
S_2(\theta)&=&S(\theta)-S_a(\theta)\rightarrow e^{i\chi
_{sG}(\theta)} \, ,  \nonumber
\end{eqnarray}
with the overall factor
\begin{equation}
e^{i\chi_{sG}(\theta)}={\mbox {exp}}\left [ i \int _{0}^{\infty}
\frac {dk}{k} \frac {\sinh k \left ( \frac
{\pi^2}{4\eta}-\pi\right)}{\sinh k \left ( \frac
{\pi^2}{4\eta}-\frac {\pi}{2}\right)\cosh \frac {\pi k}{2}}\sin
k\theta \right ] \, ,
\end{equation}
describing the sine-Gordon scattering
soliton-soliton$\rightarrow$soliton-soliton \cite{ZAM}.

\subsection{Attractive regime: $\frac {\bf K}{2}< \eta <{\bf K}$}

As anticipated, it is really worth to complete the picture by
studying the scattering theory in the attractive regime ${\bf
K}/2< \eta <{\bf K}$: here, soliton and antisoliton excitations do
not enjoy apart existences only, but also bound states. And we
fancy to call these states (elliptic) breathers in analogy with
the sine-Gordon locution. Although such bound states -- whose
number depends indeed on the value of $\eta$ -- shall have
different scattering factors, for simplicity's sake we restrict
our attention to the lightest one. In fact, we may expect that all
these scattering processes should be somehow described by the
algebraic structure of our simplest case. At any rate, their
account would deserve a separate work \cite{35}.

Let an asymptotic state of the lightest breather be created with
rapidity $\theta_i$, $i=1,2$, by a Zamolodchikov-Faddeev generator
$B(\theta_i)$ with exchange relation given by a two-particle
amplitude $S_B(\theta _1-\theta _2)$:
\begin{equation}
B(\theta_1)B(\theta _2)=S_B(\theta _1-\theta
_2)B(\theta_2)B(\theta _1) \, . \label{BB}
\end{equation}
In our set-up $S_B(\theta_1-\theta_2)$ derives from considering
the thermodynamic limit of the NLIE describing the excitation of
two lightest (elliptic) breathers. This configuration with two
breathers $B(\theta_i)$ with rapidities $\theta_i$ corresponds to
adding up, to a sea of real roots, two self-conjugate roots with
real part $\theta_i$ respectively, and imaginary part ${{\bf
K}^2}/\eta$ (limiting value). Then, the NLIE (\ref {ddv3}) takes
on the form (discarding the convolution term as usual)
\begin{equation}
Z(\tilde \theta)=NF(\tilde \theta)-\chi _{II}\left (\tilde \theta
-\theta _1 -i\frac {{\bf K}^2}{\eta } \right ) -\chi _{II}\left
(\tilde \theta -\theta _2 -i\frac {{\bf K}^2}{\eta } \right ) \, ,
\label {secondZ}
\end{equation}
when $|{\mbox {Im}}\tilde \theta |<2{{\bf K}^2}/\eta - 2{\bf K}$
is in the first analyticity strip. Nevertheless, the imaginary
part of a self-conjugate root $\frac {{\bf K}^2}{\eta }>2{{\bf
K}^2}/\eta - 2{\bf K}$ turns out to be outside the first
analyticity strip and therefore we had to use the second
determination of the function $\chi (\theta)$, i.e.
\begin{equation}
\chi_{II}(\theta)=\chi (\theta)-\chi(\theta-2i({\bf K}^2/\eta-{\bf
K}){\mbox {sgn}}{\mbox {Im}}\theta) \, , \label {secondchi}
\end{equation}
which holds as the imaginary part $|{\mbox {Im}}\theta |>2{\bf
K}^2/\eta-2{\bf K}$ lies outside the first analyticity strip.
Implementing now the definition (\ref{chi}), it follows a compact
formula for the functions $\chi _{II}(\theta)$ in (\ref{secondZ})
\begin{equation}
\chi_{II}(\theta)=i \ln \left [ \frac {(x^{-1}P^{-1}Q^2;Q^4)(
xQ^2;Q^4)(x^{-1}Q^2;Q^4)(xPQ^2;Q^4)} {(x^{-1};Q^4)(x
P;Q^4)(x^{-1}P^{-1}Q^4;Q^4)(xQ^4;Q^4)} \right ] \, ,
\end{equation}
where $P$ and $Q$ are defined in (\ref{QP}) and
\begin{equation}
x = {\mbox {exp}}\left (\frac {2i\pi \eta \theta}{\bf {K} \bf
{K}'} \right ) \, .
\end{equation}
As second and last step, we compute $Z(\tilde \theta)$ at one of
the self-conjugate roots, for instance
\begin{equation}
\tilde \theta=\theta_1 +i\frac {{\bf K}^2}{\eta }   \, ,
\label{selfroot}
\end{equation}
and expect that the terms involving $\chi _{II}$ furnish $i \ln
S_{B}(\theta _1-\theta _2)$. And again the imaginary part of a
self-conjugate root (\ref{selfroot}) $\frac {{\bf K}^2}{\eta
}>2{{\bf K}^2}/\eta - 2{\bf K}$ lies outside the first analyticity
strip of $Z(\tilde \theta)$ itself: therefore we need to consider
its second determination as well. In other words, we must
introduce the second determination of $\chi_{II}(\theta)$,
$\chi_{II}(\theta)_{II}$:
\begin{equation}
Z_{II}\left(\theta_1 +i\frac {{\bf K}^2}{\eta } \right )= NF\left(
\theta_1 +i\frac {{\bf K}^2}{\eta } \right )_{II}-\chi _{II}(
0)_{II} -\chi _{II}(\theta _1-\theta _2)_{II} \, . \label{secsecZ}
\end{equation}
We can calculate it by reiterating (\ref {secondchi}) with now
${\mbox {Im}} \theta> 0$ and simply obtain
\begin{equation}
\chi _{II}(\theta)_{II}=i \ln \left [ -x \frac
{(x^{-1}P^{-1}Q^2;Q^4)(x^{-1}P;Q^4)(x P^{-1}Q^4;Q^4)(x
PQ^2;Q^4)}{(xP^{-1}Q^2;Q^4)(x P;Q^4)(x^{-1}P^{-1}Q^4;Q^4)(x^{-1}
PQ^2;Q^4)} \right ] \, ,
\end{equation}
where we have eventually identified
\begin{equation}
\theta=\theta_1-\theta_2, \,\,\, x={\mbox {exp}}\left (\frac
{2i\pi \eta \theta}{\bf {K} \bf {K}'} \right ) \, .
\end{equation}
In conclusion, the scattering factor between the lightest elliptic
breathers reads in this XYZ notation
\begin{eqnarray}
&&S_{B}(\theta)=-{\mbox {exp}}\left [i\chi _{II}(0)_{II} \right]
{\mbox {exp}}\left [i\chi _{II}(\theta)_{II} \right]=\nonumber \\
&=&-\frac {1}{x}\frac
{(xP^{-1}Q^2;Q^4)(xP;Q^4)(x^{-1}P^{-1}Q^4;Q^4)(x^{-1}PQ^2;Q^4)}{(x^{
-1}P^{-1}Q^2;Q^4)(x^{-1}P;Q^4)(xP^{-1}Q^4;Q^4)(xPQ^2;Q^4)} \, ,
\end{eqnarray}
with the above $\theta$ and $x$. An overall minus sign has been
permitted thanks to the definition of $\chi_{II}$ up to a constant
and in order to reproduce $S_B(0)=-1$ as in the sine-Gordon field
theory.

Despite the cumbersome calculation and the different origin, this
scattering factor coincides exactly with the structure function of
the Deformed Virasoro Algebra ${\cal V}ir_{p_v,q_v}$ introduced by
Shiraishi et al. \cite{SKAO} . The Deformed Virasoro Algebra (DVA)
is an associative algebra generated by the modes $T_n$ of the
current $T(z)=\sum _n T_n z^{-n}$, satisfying the relation
\begin{equation}
f(w/z)T(z)T(w)-f(z/w)T(w)T(z)=-\frac
{(1-q_v)(1-p_vq_v^{-1})}{1-p_v}\left [\delta \left (\frac
{p_vw}{z} \right )- \delta \left (\frac {w}{p_vz} \right ) \right
] \, , \label{qpVir}
\end{equation}
where $p_v$ and $q_v$ are complex parameters and
\begin{equation}
f(x_v)={\mbox {exp}}\left [\sum _{n=1}^{\infty}\frac
{(1-q_v^n)(1-q_v^{-n}p_v^n)}{1+p_v^n}\frac {x_v^n}{n}\right] \, .
\end{equation}
Equality (\ref{qpVir}) is to be interpreted as an equality between
formal power series, but, as shown in Appendix B.1, $f(x_v)$ can
be analytically continued to the whole complex plane and this
allows us to recast relation (\ref{qpVir}) in the braiding form
\begin{equation}
T(z)T(w)=Y(z/w)T(w)T(z) \, ,
\end{equation}
where
\begin{equation}
Y(x_v)=-\frac {1}{x_v}\frac
{(x_vq_v^{-1}p_v;p_v^2)(x_vq_v;p_v^2)(x_v^{-1}q_v^{-1}p_v^2;
p_v^2)(x_v^{-1}q_vp_v;p_v^2)}{(x_v^{-1}q_v^{-1}p_v;p_v^2)(x_v^{-1}q_v;
p_v^2)(x_vq_v^{-1}p_v^2;p_v^2)(x_vq_vp_v;p_v^2)}
\label{DVA1}
\end{equation}
is the structure function of the DVA. While writing this formula,
we have borne in mind the convenient infinite product notation
\begin{equation}
(x;a)=\prod _{s=0}^{\infty}(1-xa^s) \, , \nonumber
\end{equation}
introduced in (\ref{infprod}). Eventually, if we suppose the
mapping between spin chain variables and algebra parameters
\begin{equation}
x_v=x^{-1} = {\mbox {exp}}\left (-\frac {2i\pi \eta \theta}{\bf
{K} \bf {K}'} \right ) \, , \quad q_v= Q^2 P ={\mbox {exp}}\left
(-\frac {4\pi \bf {K}} {\bf {K}'} \right ) \, , \quad p_v=Q^2
={\mbox {exp}}\left (-\frac {4\eta \pi } {\bf {K}'} \right )\, ,
\label{DVAnota}
\end{equation}
we can see immediately that
\begin{equation}
S_B(\theta)=Y(x_v)\, .
   \label{DVAequ}
\end{equation}
Therefore, we have proved that the Deformed Virasoro Algebra
defines a current, $T(z)$, which closes the Zamolodchikov-Faddeev
algebra for the fundamental scalar excitation of the XYZ model.

\medskip

It is possible to rewrite $S_B(\theta)$ in a different form, which
is useful if we want to compare this factor to analogous ones
proposed in the literature. We follow the idea of Lukyanov
\cite{LUK}, write (\ref{DVAequ},\ref{DVA1}) in terms of
theta-functions, perform a modular transformation on the elliptic
nome and eventually obtain (details in Appendix B.2)
\begin{equation}
S_B(\theta)=\frac {\theta _{11}\left (\frac {i\theta}{4{\bf
K}}+\frac {{\bf K}}{2\eta};\frac {i {\bf K}'}{4\eta}\right) \theta
_{10}\left (\frac {i\theta}{4{\bf K}}-\frac {{\bf K}}{2\eta};\frac
{i {\bf K}'}{4\eta}\right)}
      {\theta _{11}\left (\frac {i\theta}{4{\bf K}}-\frac {{\bf
K}}{2\eta};\frac {i {\bf K}'}{4\eta}\right) \theta _{10}\left
(\frac {i\theta}{4{\bf K}}+\frac {{\bf K}}{2\eta};\frac {i {\bf
K}'}{4\eta}\right)} \label {DVAluk} \, .
\end{equation}
This is  a useful form as it allows us to perform easily the limit
${\bf K}'\rightarrow \infty$ which describes -- after shifting
(Section 5) -- the sine-Gordon model. Indeed, we obtain the first
breather scattering factor \cite{ZAM}
\begin{equation}
\lim _{{\bf K}'\rightarrow \infty}S_B(\theta)= \frac {\sinh \theta
+ i \sin \pi \xi}{\sinh \theta - i \sin \pi \xi } \, ,
\end{equation}
with the usual parameter
\begin{equation}
\xi=\frac {\pi}{2\eta}-1 \, .
\end{equation}
Moreover, it follows from (\ref{etabiden}) that $\xi$ carries on
the dependence on the coupling constant $b$ of the sine-Gordon
Lagrangian:
\begin{equation}
\xi = \frac {b^2}{8\pi - b^2} \, . \label{xib}
\end{equation}

It is an easy job, now, to relate (\ref{DVAluk}) to the scattering
factor $S_{MP}(\beta)$ proposed by Mussardo and Penati in
\cite{MP}. We start from the form (6) in \cite{MP} for
$S_{MP}(\beta)$ and after some manipulations (details in Appendix
B.3), we can conclude that this expression (6) equals our
(\ref{DVAluk}), namely $S_{MP}(\beta)= S_B(\theta)$, provided
these identifications are taken into account (parameters of
\cite{MP} are on the l.h.s.)
\begin{equation}
T_{MP}=\frac {\pi{\bf K}'}{2\eta} \, , \quad a_{MP}=\frac {{\bf
K}}{\eta }-1 \, , \quad \beta_{MP}=-\frac {\pi \theta}{2{\bf K}}
\, . \label {Sbrea}
\end{equation}
As an obvious but intriguing byproduct we want to underline the
coincidence of the Mussardo-Penati's scattering factor and the
structure function of the DVA.

\medskip

{\bf Remark}: The parameters $x_v$, $q_v$ and $p_v$ of the
attractive regime (and of the DVA structure function) inherit a
simple expression in terms of the parameters $x_e$, $q_e$ and
$p_e$ of the repulsive regime (and of the R-matrix of the elliptic
algebra ${\cal A}_{q_e,p_e}(\widehat {sl}(2)_c)$):
\begin{equation}
x_v=x_e^2 \, , \quad p_v=q_e^2 \, , \quad q_v=p_e  \, .
\label{qPrel}
\end{equation}
Although in a different landscape, this seems to be exactly the
connection between $Vir_{p_v,q_v}$ and ${\cal
A}_{q_e,p_e}(\widehat {sl}(2)_c)$ proved in \cite{JS} \footnote{As
a conjecture this liaison is already in \cite{XMAS} and somehow in
\cite{AFRS}.}.

\section{Some conclusions, many prospects}
Starting from the (elliptic) Bethe Ansatz equations of the spin
$1/2$-XYZ chain on a circumference (cfr. for instance \cite{TF}),
we have written a Non-Linear Integral Equation describing
describing a generic state (either the vacuum or an excited state)
of the model in the disordered regime. Maintaining the size
finite, we have studied two different limits: the usual
trigonometric limit which gives the spin 1/2-XXZ chain and a
double scaling limit which turns out to describe the sine-Gordon
field theory on a cylinder. The latter furnishes the
generalization of the infinite length scaling limit of \cite{LUT}
and \cite{JKM} in case of finite volume. Moreover, it has
suggested us the heuristic idea that sending to infinity the size
of the $1/2$-XYZ would infer an elliptic scattering theory, which
naturally inherits an elliptic deformation of all the sine-Gordon
structures. In fact, any elementary excitation on the vacuum in
the repulsive regime can be completely characterised, through a
NLIE, by new terms which give rise to the corresponding scattering
amplitude. Therefore, the elliptic deformation of the
soliton/antisoliton sine-Gordon S-matrix derives from the finite
size procedure and the re-expression of this matrix as the Baxter
elliptic R-matrix proves what is called in \cite{FIJKMY} the
Smirnov's conjecture, a remarkable connection between
representation theory of an elliptic algebra and a scattering
S-matrix. Very likely the conjecture may also be extended to more
complicated elliptic algebras, being its formulation purely
algebraic.

Moreover, we have studied the lightest bound state, i.e. the
elliptic deformation of the first sine-Gordon breather, and found
that its scattering factor coincides with a proposal by Lukyanov
\cite{LUK} and, in a different context, by Mussardo-Penati
\cite{MP} \footnote{The coincidence of these two was apparently
unnoticed beforehand.}, provided a mapping between each set of
coupling constants and scattering rapidities is given. Actually,
the most important result of this calculation is that the
scattering factor coming from the spin chain shows manifestly its
identity with the braiding factor of the Deformed Virasoro Algebra
by Shiraishi-Kubo-Awata-Odake \cite{SKAO} and let us suppose that
each heavier breather may produce a braiding algebra with a
similar structure (possibly like in \cite{AFRS}, but without one
of the squares in the structure function). In this respect, we
would like to clarify this point in a ongoing publication
\cite{35}, since we reckon the basic DVA of \cite{SKAO} as
responsible somehow for the algebraic structure and the mass
quantisation of the other breathers as well. This conjecture
relies upon the identification of the first breather as the
fundamental scalar particle and upon the vertex operator
construction of DVA given in \cite{AKMOS}. We would like to
conclude this part about the relevance of the DVA by highlighting
how its field theory limit in sine-Gordon should be the quantum
version of the non-local symmetry geometrically constructed in
\cite{FS}. But the field theory inheritance really needs an ad hoc
analysis in a separate paper, though the scenario is becoming
clearer thanks to these spin chain developments. Nevertheless, it
is also worth analysing the whole information obtainable on the
lattice system through at least two routes: on the one hand,
pretty much is now known about the representation theory of the
Deformed Virasoro Algebra (cfr. \cite{XMAS} as a review work); on
the other hand, the form factors postulates authorised Mussardo
and Penati to work out an entire series of form factors of not
better identified ''fields'' \cite{MP} (where also some
speculations on the nature of the fields is brought forward in the
light of conformal fields structure). By now, a more precise
reading of these form factors becomes possible and a field theory
interpretation plausible.

In conclusion, a field theory description of the spin chain in the
thermodynamic limit, also conveying and re-interpreting previous
results, shall be pursued by means of fermionisation and
renormalisation group techniques and would contribute to the
clarity of the landscape. Towards this aim, valuable results from
\cite{LS} might be efficiently extracted.

\vspace {1cm} {\bf Acknowledgements} - Discussions with R. Sasaki,
P. Sorba, H. Konno and J. Suzuki are kindly acknowledged. D.F.
thanks Leverhulme Trust for grant F/00224/G and EC FP5 Network
EUCLID (contract number HPRN-CT-2002-00325) for partial financial
support. M.R. thanks JSPS for the Postdoctoral Fellowship
(Short-term) for North American and European Researchers PE03003
and for the Invitation Fellowship for Research in Japan
(Long-term) L04716 and the Department of Mathematics of the
University of York for warm and kind hospitality.

\renewcommand{\thesection}{\Alph{section}}

\section*{Appendix A}
\setcounter{section}{1} \setcounter{equation}{0} In this appendix
we want to compute the Fourier coefficient of the function
\begin{equation}
\frac {d}{dx}\phi (x+i\alpha, \eta)  \, , \quad \alpha \in
{\mathbb R} \, .
\end{equation}
  From the definition (\ref {phi}) of $\phi$ it follows that we have to
consider the expression
\begin{eqnarray}
\frac {d}{dx}\phi (x+i\alpha, \eta)&=&i \frac {d}{dx} \ln
H(\eta+\alpha
-ix)-i \frac {d}{dx} \ln H(\eta -\alpha+ix)+\nonumber \\
&+&i \frac {d}{dx} \ln \Theta (\eta +\alpha-ix)-i \frac {d}{dx}
\ln \Theta (\eta -\alpha+ix) \, .
\end{eqnarray}
Using formula (8.199.1) of \cite{GRA} we have:
\begin{eqnarray}
&&i \frac {d}{dx} \ln H(\eta -ix)=\frac {\pi}{2 {\bf K}}\left
[\cot \frac {\pi (\eta -ix)}{2{\bf K}}-2i\sum _{n=1}^\infty \frac
{e^{\frac {i\pi}{\bf K}(\eta  -ix)}-e^{-\frac {i\pi}{\bf K}(\eta
-ix)}}{e^{\frac {2\pi {\bf K}' n}{\bf K}}-e^{\frac {i\pi}{\bf
K}(\eta -ix)}-e^{-\frac {i\pi}{\bf K}(\eta  -ix)}+e^{-\frac {2\pi
{\bf K}' n}{\bf K} }}\right]
\nonumber \\
&=&\frac {\pi}{2 {\bf K}}\left [\cot \frac {\pi (\eta -ix)}{2{\bf
K}}-2i\sum _{n=1}^\infty \frac {e^{-\frac {2\pi {\bf K}' n}{\bf
K}}\left (e^{\frac {i\pi \eta +\pi x}{\bf K}}-e^{-\frac {i\pi \eta
+\pi x}{\bf K}}\right) }{\left (1-e^{\frac {i\pi \eta +\pi x-2\pi
{\bf K}' n}{\bf K}}\right)\left (1-e^{-\frac {i\pi \eta +\pi
x+2\pi {\bf K}' n}{\bf K}}\right)}\right] \, .
\end{eqnarray}
Since $-\frac {{\bf K}'}{2}<x<\frac {{\bf K}'}{2}$, we can express
the denominators as power series:
\begin{eqnarray}
&&i \frac {d}{dx} \ln H(\eta  -ix)=\frac {\pi}{2 {\bf K}}\left
[\cot \frac {\pi (\eta  -ix)}{2{\bf K}}-2i\sum _{n=1}^\infty
\sum_{j,l=0}^\infty e^{-\frac {2\pi {\bf K}' n}{\bf K}}\left
(e^{\frac {i\pi \eta +\pi x}{\bf K}}-e^{-\frac {i\pi \eta +\pi
x}{\bf K}}\right)
\cdot \right. \nonumber \\
&\cdot & \left. e^{j\frac {i\pi \eta +\pi x-2\pi {\bf K}' n}{\bf
K}}e^{-l\frac {i\pi \eta +\pi x+2\pi {\bf K}' n}{\bf K}}\right] =
\nonumber \\
&=&\frac {\pi}{2 {\bf K}}\Bigl [\cot \frac {\pi (\eta -ix)}{2{\bf
K}}-2i\sum _{n=1}^\infty \sum_{j,l=0}^\infty \left (e^{\frac {i\pi
\eta +\pi x}{\bf K}(1+j-l)-\frac {2\pi {\bf K}' n}{\bf
K}(1+j+l)}-e^{-\frac {i\pi \eta +\pi x}{\bf K}(1+l-j)-\frac {2\pi
{\bf K}' n}{\bf K}(1+j+l)} \right)
\Bigr ] \nonumber \\
&=&\frac {\pi}{2 {\bf K}}\Bigl \{ \cot \frac {\pi (\eta
-ix)}{2{\bf K}}-2i\sum _{n=1}^\infty \sum _{S=0}^\infty e^{ -\frac
{2\pi {\bf K}' n}{\bf K}(1+S)} \sum _{D=-S/2}^{S/2}\left [
e^{\frac {i\pi \eta  +\pi x}{\bf K}(1+2D)}-e^{-\frac {i\pi \eta
+\pi x}{\bf K}(1-2D)} \right ] \Bigr \}
= \nonumber \\
&=&\frac {\pi}{2 {\bf K}}\Bigl \{ \cot \frac {\pi (\eta
-ix)}{2{\bf K}}+2i\sum _{n=1}^\infty \sum _{S=0}^\infty e^{ -\frac
{2\pi {\bf K}' n}{\bf K}(1+S)} \left [ e^{-\frac {i\pi \eta  +\pi
x}{\bf K}(S+1)}- e^{\frac {i\pi \eta  +\pi
x}{\bf K}(S+1)}\right ] \Bigr \} = \nonumber \\
&=&\frac {\pi}{2 {\bf K}}\Bigl \{ \cot \frac {\pi (\eta
-ix)}{2{\bf K}}+2i \sum _{S=1}^\infty \frac {e^{ -\frac {2\pi {\bf
K}' }{\bf K}S}}{1-e^{ -\frac {2\pi {\bf K}' }{\bf K}S}} \left [
e^{-\frac {i\pi \eta  +\pi x}{\bf K}S}- e^{\frac {i\pi \eta +\pi
x}{\bf K}S}\right ] \Bigr \} \, .\nonumber
\end{eqnarray}
Now, we are ready to perform the integration
\begin{equation}
\int _{-\frac {{\bf K}'}{2}}^{\frac {{\bf K}'}{2}} dx \left [ i
\frac {d}{dx} \ln H(\eta +\alpha-ix)-i \frac {d}{dx} \ln H(\eta
-\alpha+ix)\right] e^{2i \frac {n \pi x}{{\bf K}'}} \, . \label
{Hcontr}
\end{equation}
About the terms depending on the cotangents we have that
\begin{eqnarray}
&&\frac {\pi}{2 {\bf K}}\int _{-\frac {{\bf K}'}{2}}^{\frac {{\bf
K}'}{2}} dx \left [ \cot \frac {\pi (\eta+\alpha -ix)}{2{\bf K}}+
\cot \frac {\pi (\eta-\alpha +ix)}{2{\bf K}}\right] e^{2i \frac {n
\pi
x}{{\bf K}'}}= \nonumber \\
&=&\frac {i\pi}{ {\bf K}}\sum _{S=1}^{\infty}\left ( e^{-\frac
{i\pi S (\eta+\alpha)}{\bf K}}- e^{\frac {i\pi S
(\eta-\alpha)}{\bf K}}\right) \frac {(-1)^ne^{-\frac {\pi S {\bf
K}'}{2\bf K}}-1}{\frac {2in\pi}{{\bf
K}'}-\frac {\pi S}{\bf K}}+\nonumber \\
&+&\frac {i\pi}{ {\bf K}}\sum _{S=1}^{\infty}\left ( e^{\frac
{i\pi S (\eta+\alpha)}{\bf K}}- e^{-\frac {i\pi S
(\eta-\alpha)}{\bf K}}\right) \frac {(-1)^ne^{-\frac {\pi S {\bf
K}'}{2\bf K}}-1}{\frac {2in\pi}{{\bf K}'}+\frac {\pi S}{\bf K}} \,
. \label{cotg}
\end{eqnarray}
On the other hand, the remaining terms yield
\begin{eqnarray}
&&\frac {i\pi}{ {\bf K}}\int _{-\frac {{\bf K}'}{2}}^{\frac {{\bf
K}'}{2}} dx e^{2i \frac {n \pi x}{{\bf K}'}}\sum _{S=1}^\infty
\frac {e^{ -\frac {2\pi {\bf K}'}{\bf K}S}}{1-e^{ -\frac {2\pi
{\bf K}'}{\bf K}S}} \left [ e^{-\frac {i\pi ( \eta+\alpha) +\pi
x}{\bf K}S}- e^{\frac {i\pi (\eta+\alpha) +\pi x}{\bf K}S} +
(x,\alpha \rightarrow
-x,-\alpha) \right ]  = \nonumber \\
&=& \frac {i\pi}{ {\bf K}}\sum _{S=1}^\infty \frac {e^{ -\frac
{2\pi {\bf K}'}{\bf K}S}}{1-e^{ -\frac {2\pi {\bf K}'}{\bf
K}S}}(-1)^n \left ( e^{-\frac {i\pi (\eta +\alpha) S}{\bf K}}\frac
{e^{-\frac {\pi {\bf K}' S}{2\bf K}}-e^{\frac {\pi {\bf K}'
S}{2\bf K}}}{\frac {2in\pi}{{\bf K}'}-\frac {\pi S}{\bf K}}-
e^{\frac {i\pi (\eta +\alpha) S}{\bf K}}\frac {e^{\frac {\pi {\bf
K}' S}{2\bf K}}-e^{-\frac {\pi {\bf K}' S}{2\bf K}}}{\frac
{2in\pi}{{\bf K}'}+\frac {\pi S}{\bf K}} - \right.
\nonumber \\
&& \left. - e^{-\frac {i\pi (\eta -\alpha) S}{\bf K}}\frac
{e^{-\frac {\pi {\bf K}' S}{2\bf K}}-e^{\frac {\pi {\bf K}'
S}{2\bf K}}}{\frac {2in\pi}{{\bf K}'}+\frac {\pi S}{\bf
K}}+e^{\frac {i\pi (\eta -\alpha)S}{\bf K}}\frac {e^{\frac {\pi
{\bf K}' S}{2\bf K}}-e^{-\frac {\pi {\bf K}' S}{2\bf K}}}{\frac
{2in\pi}{{\bf K}'}-\frac {\pi S}{\bf K}}\right ) \, .   \label
{Hfunct}
\end{eqnarray}
Summing these two expressions we get the result for (\ref
{Hcontr}). However, before doing such a sum, it is convenient to
compute the contribution coming from the $\Theta $ function:
\begin{equation}
\int _{-\frac {{\bf K}'}{2}}^{\frac {{\bf K}'}{2}} dx \left [ i
\frac {d}{dx} \ln \Theta (\eta +\alpha-ix)-i \frac {d}{dx} \ln
\Theta (\eta -\alpha+ix)\right] e^{2i \frac {n \pi x}{{\bf K}'}}
\, . \label{tetacontri}
\end{equation}
Using formula 8.199.4 of \cite {GRA} we get
\begin{eqnarray}
&& i \frac {d}{dx} \ln \Theta (\eta -ix)=\frac {2\pi}{\bf K}\sum
_{n=1}^\infty \frac {\sin \frac {\pi} {\bf K}(\eta -ix)}{e^{\frac
{\pi {\bf K}' (2n-1)}{\bf K}}-e^{\frac {i\pi}{\bf K}(\eta
-ix)}-e^{-\frac {i\pi}{\bf
K}(\eta -ix)}+e^{-\frac {\pi {\bf K}' (2n-1)}{\bf K} } }=\nonumber \\
&=&\frac {2\pi}{\bf K} \sum _{n=1}^\infty \frac {\sin \frac {\pi}
{\bf K}(\eta -ix)e^{-\frac {\pi {\bf K}'  (2n-1)}{\bf K}}}{\left
[1-e^{-\frac {\pi {\bf K}' (2n-1)}{\bf K}}e^{\frac {i\pi}{\bf
K}(\eta -ix)}][1-e^{-\frac {\pi {\bf K}' (2n-1)}{\bf K}}e^{-\frac
{i\pi}{\bf K}(\eta -ix)}\right ]} \, .
\end{eqnarray}
Since $-\frac {{\bf K}'}{2}<x<\frac {{\bf K}'}{2}$ we can express
as power series the denominators:
\begin{eqnarray}
&& i \frac {d}{dx} \ln \Theta (\eta -ix)=\frac {2\pi}{\bf K}\sin
\frac {\pi} {\bf K}(\eta -ix) \sum _{n=1}^\infty
e^{-\frac {\pi {\bf K}' (2n-1)}{\bf K}} \cdot \nonumber \\
&\cdot & \sum _{j,l=0}^\infty e^{-\frac {\pi {\bf K}' (2n-1)}{\bf
K}l+\frac {i\pi} {\bf K}(\eta -ix)l } e^{-\frac {\pi {\bf K}'
(2n-1)}{\bf K}j-\frac {i\pi} {\bf K}(\eta
-ix)j }=\nonumber \\
&=& \frac {2\pi}{\bf K}\sin \frac {\pi} {\bf K}(\eta -ix) \sum
_{n=1}^\infty  \sum _{j,l=0}^\infty e^{-\frac {\pi {\bf K}' }{\bf
K}(2n-1)(j+l+1)}e^{\frac {i\pi} {\bf K}(\eta -ix)(l-j)}=\nonumber \\
&=&\frac {2\pi}{\bf K}\sin \frac {\pi} {\bf K}(\eta -ix) \sum
_{n=1}^\infty  \sum _{S=0}^\infty \sum _{D=-S/2}^{S/2} e^{-\frac
{\pi {\bf K}' }{\bf K}(2n-1)(S+1)}e^{\frac {i\pi} {\bf K}(\eta
-ix)2D}=\nonumber \\
&=&\frac {i\pi}{\bf K} \sum _{n=1}^\infty \sum _{S=0}^\infty
e^{-\frac {\pi {\bf K}' }{\bf K}(2n-1)(S+1)}\left [ e^{-\frac
{i\pi} {\bf K}(\eta -ix)(S+1)}-e^{\frac {i\pi} {\bf K}(\eta
-ix)(S+1)} \right ] =\nonumber
\\
&=& \frac {i\pi}{\bf K}\sum _{S=1}^\infty \frac {e^{\frac {\pi
{\bf K}' }{\bf K}S}}{e^{2\frac {\pi {\bf K}' }{\bf K}S}-1}\left [
e^{-\frac {i\pi} {\bf K}(\eta -ix)S}-e^{\frac {i\pi} {\bf K}(\eta
-ix)S} \right ] \, . \label {tetaexp}
\end{eqnarray}
With the help of (\ref {tetaexp}) we can perform the integration
involved in (\ref {tetacontri}):
\begin{eqnarray}
&&\int _{-\frac {{\bf K}'}{2}}^{\frac {{\bf K}'}{2}} dx \left [ i
\frac {d}{dx} \ln \Theta (\eta+\alpha -ix)-i \frac {d}{dx} \ln
\Theta (\eta -\alpha+ix)\right]
e^{2i \frac {n \pi x}{{\bf K}'}} = \nonumber \\
&=& \frac {i\pi}{ {\bf K}}\sum _{S=1}^\infty \frac {1}{e^{ \frac
{2\pi {\bf K}'}{\bf K}S}-1}e^{ \frac {\pi {\bf K}'}{\bf K}S}(-1)^n
\left ( e^{-\frac {i\pi (\eta +\alpha)S}{\bf K}}\frac {e^{-\frac
{\pi {\bf K}' S}{2\bf K}}-e^{\frac {\pi {\bf K}' S}{2\bf
K}}}{\frac {2in\pi}{{\bf K}'}-\frac {\pi S}{\bf K}}- e^{\frac
{i\pi (\eta +\alpha)S}{\bf K}}\frac {e^{\frac {\pi {\bf K}'
S}{2\bf K}}-e^{-\frac {\pi {\bf K}' S}{2\bf K}}}{\frac
{2in\pi}{{\bf K}'}+\frac {\pi S}{\bf K}} - \right.
\nonumber \\
&& \left. - e^{-\frac {i\pi (\eta -\alpha)S}{\bf K}}\frac
{e^{-\frac {\pi {\bf K}' S}{2\bf K}}-e^{\frac {\pi {\bf K}'
S}{2\bf K}}}{\frac {2in\pi}{{\bf K}'}+\frac {\pi S}{\bf
K}}+e^{\frac {i\pi (\eta -\alpha)S}{\bf K}}\frac {e^{\frac {\pi
{\bf K}' S}{2\bf K}}-e^{-\frac {\pi {\bf K}' S}{2\bf K}}}{\frac
{2in\pi}{{\bf K}'}-\frac {\pi S}{\bf K}}\right ) \label {teta} \,.
\end{eqnarray}
Now, summing (\ref {Hfunct}) and (\ref {teta}), we get
\begin{equation}
-\frac {i\pi}{ {\bf K}}\sum _{S=1}^{\infty}\left [ \left (
e^{-\frac {i\pi S (\eta+\alpha)}{\bf K}}- e^{\frac {i\pi S
(\eta-\alpha)}{\bf K}}\right) \frac {(-1)^ne^{-\frac {\pi S {\bf
K}'}{2\bf K}}}{\frac {2in\pi}{{\bf K}'}-\frac {\pi S}{\bf
K}}-\left ( e^{\frac {i\pi S (\eta+\alpha)}{\bf K}}- e^{-\frac
{i\pi S (\eta-\alpha)}{\bf K}}\right) \frac {(-1)^ne^{-\frac {\pi
S {\bf K}'}{2\bf K}}}{\frac {2in\pi}{{\bf K}'}+\frac {\pi S}{\bf
K}} \right] \, . \label{hteta}
\end{equation}
We notice that this expression exactly cancels the term in (\ref
{cotg}) proportional to $(-1)^n$. Therefore, we get the following
result for the Fourier coefficient of $\frac {d}{dx}\phi
(x+i\alpha, \eta), \, \alpha \in {\mathbb R}$:
\begin{equation}
\int _{-\frac {{\bf K}'}{2}}^{\frac {{\bf K}'}{2}} dx \frac
{d}{dx}\phi (x+i\alpha, \eta)  e^{2i \frac {n \pi x}{{\bf
K}'}}=\frac {i\pi}{ {\bf K}}\sum _{S=1}^{\infty}\left [ \frac
{e^{\frac {i\pi S (\eta-\alpha)}{\bf K}}- e^{-\frac {i\pi S
(\eta+\alpha)}{\bf K}}}{\frac {2in\pi}{{\bf K}'}-\frac {\pi S}{\bf
K}}+\frac {e^{-\frac {i\pi S (\eta-\alpha)}{\bf K}}- e^{\frac
{i\pi S (\eta+\alpha)}{\bf K}}}{\frac {2in\pi}{{\bf K}'}+\frac
{\pi S}{\bf K}} \right]  \, . \label{A12}
\end{equation}
The sum over $S$ can be performed using formul{\ae} 1.445.1,2 of
\cite {GRA}. The final result depends on the range of values of
$\alpha$ and $\eta$. We remark that we can remove the restriction
to $\alpha$ real, since the imaginary part of $\alpha$ can be
easily implemented in the final formula, its effect being a phase.
Because of the periodicity property  $\phi '(x+2i{\bf K})=\phi
'(x)$, we can restrict $\alpha$ to the interval $-\eta <{\mbox
{Re}}\alpha <2{\bf K} -\eta$. We remember also that $0<\eta<{\bf
K}$ (\ref {regime}). We get that, for $\alpha \in {\mathbb C}$:
\begin{eqnarray}
      \int _{-\frac {{\bf K}'}{2}}^{\frac {{\bf K}'}{2}} dx \frac
{d}{dx}\phi (x+i\alpha, \eta)  e^{2i \frac {n \pi x}{{\bf
K}'}}&=&2\pi \frac {\sinh \frac {2n({\bf K}-\eta)\pi}{{\bf
K}'}}{\sinh \frac {2n{\bf K}\pi}{{\bf K}'}} e^{\frac {2n\pi
\alpha}{{\bf K}'}} \, , \quad -\eta
<{\mbox {Re}}\alpha <\eta \, , \nonumber \\
       \label {fourphiprimegen} \\
      \int _{-\frac {{\bf K}'}{2}}^{\frac {{\bf K}'}{2}} dx \frac
{d}{dx}\phi (x+i\alpha, \eta)  e^{2i \frac {n \pi x}{{\bf
K}'}}&=&-2\pi \frac {\sinh \frac {2n\eta\pi}{{\bf K}'}}{\sinh
\frac {2n{\bf K}\pi}{{\bf K}'}} e^{-\frac {2n\pi ({\bf
K}-\alpha)}{{\bf K}'} }\, , \quad \eta <{\mbox {Re}}\alpha <2{{\bf
K}}-\eta \, . \nonumber
\end{eqnarray}
On the other hand, formula (\ref {A12}) holds also if we replace
$\eta $ with $2\eta$. Since now $0<2\eta<2{\bf K}$, in order to
compute the sums over $S$ we distinguish the following cases:

$\bullet$ $0<\eta<{\bf K}/2 $:
\begin{eqnarray}
      \int _{-\frac {{\bf K}'}{2}}^{\frac {{\bf K}'}{2}} dx \frac
{d}{dx}\phi (x+i\alpha, 2\eta)  e^{2i \frac {n \pi x}{{\bf
K}'}}&=&2\pi \frac {\sinh \frac {2n({\bf K}-2\eta)\pi}{{\bf
K}'}}{\sinh \frac {2n{\bf K}\pi}{{\bf K}'}} e^{\frac {2n\pi
\alpha}{{\bf K}'}} \, , \quad
-2\eta <{\mbox {Re}}\alpha <2\eta \, , \nonumber \\
       \label {fourphiprimegen1} \\
      \int _{-\frac {{\bf K}'}{2}}^{\frac {{\bf K}'}{2}} dx \frac
{d}{dx}\phi (x+i\alpha,2\eta)  e^{2i \frac {n \pi x}{{\bf
K}'}}&=&-2\pi \frac {\sinh \frac {4n\eta\pi}{{\bf K}'}}{\sinh
\frac {2n{\bf K}\pi}{{\bf K}'}} e^{-\frac {2n\pi ({\bf
K}-\alpha)}{{\bf K}'} }\, , \quad 2\eta <{\mbox {Re}}\alpha
<2{{\bf K}}-2\eta \, ; \nonumber
\end{eqnarray}
$\bullet$ ${\bf K}/2<\eta<{\bf K}$:
\begin{eqnarray}
\int _{-\frac {{\bf K}'}{2}}^{\frac {{\bf K}'}{2}} dx \frac
{d}{dx}\phi (x+i\alpha,2\eta)  e^{2i \frac {n \pi x}{{\bf
K}'}}&=&2\pi \frac {\sinh \frac {2n(2{\bf K}-2\eta)\pi}{{\bf
K}'}}{\sinh \frac {2n{\bf K}\pi}{{\bf K}'}} e^{\frac {2n\pi}{{\bf
K}'}(\alpha +{\bf K})} \, ,
\quad -2\eta <{\mbox {Re}}\alpha <2\eta-2{{\bf K}} \, , \nonumber \\
      \label {fourphiprimegen2} \\
      \int _{-\frac {{\bf K}'}{2}}^{\frac {{\bf K}'}{2}} dx \frac
{d}{dx}\phi (x+i\alpha, 2\eta)  e^{2i \frac {n \pi x}{{\bf
K}'}}&=&2\pi \frac {\sinh \frac {2n({\bf K}-2\eta)\pi}{{\bf
K}'}}{\sinh \frac {2n{\bf K}\pi}{{\bf K}'}} e^{\frac
{2n\pi\alpha}{{\bf K}'}} \, , \quad 2\eta-2{\bf K} <{\mbox
{Re}}\alpha <2{\bf K}-2\eta \, . \nonumber
      \end{eqnarray}

\renewcommand{\thesection}{\Alph{section}}

\section*{Appendix B: DVA as scattering algebra: a unifying outlook}
\setcounter{section}{2} \setcounter{equation}{0}

\subsection{XYZ and DVA: an equivalence}

In the definition of the Deformed Virasoro Algebra (DVA), the main
ingredient is the formal expression
\begin{equation}
f(x_v)={\mbox {exp}}\left [\sum _{n=1}^{\infty}\frac
{(1-q_v^n)(1-q_v^{-n}p_v^n)}{1+p_v^n}\frac {x_v^n}{n}\right] \, .
\label {fDVA}
\end{equation}
The series contained in (\ref {fDVA}) is convergent when $|x_v|<1$
and if, for instance, the parameters $q_v$, $p_v$ satisfy
$|p_v|<1$, $|q_v|<1$ and $|p_vq_v^{-1}|<1$. In such a case we can
sum up the series and define a meromorphic function in the complex
plane as its analytic continuation. Now, we want to determine such
a function.

We observe that, since $|p_v|<1$, it is possible to write the
denominator in (\ref{fDVA}) as a geometric series:
\begin{equation}
f(x_v)={\mbox {exp}}\left [\sum
_{n=1}^{\infty}(1-q_v^n)(1-q_v^{-n}p_v^n) \sum _{s=0}^{\infty}
(-1)^s p_v^{ns}\frac {x_v^n}{n}\right] \, . \label {fDVA2}
\end{equation}
Since $|q_v|<1$ and $|p_vq_v^{-1}|<1$, we can exchange the series
in (\ref {fDVA2}) and perform the summation over $n$ in a suitable
range of values for $x_v$ by using the formula
\begin{equation}
\sum _{n=1}^{\infty}\frac {z^n}{n}=\log \frac {1}{1-z} \, , \quad
|z|<1 \, .
\end{equation}
So we obtain
\begin{equation}
f(x_v)={\mbox {exp}}\left [ \sum _{s=0}^{\infty}(-1)^s \left (
\log \frac {1}{1-x_vp_v^s}-\log \frac {1}{1-x_vq_v^{-1}p_v^{1+s}}-
\log \frac {1}{1-x_vq_vp_v^s}+\log \frac {1}{1-x_vp_v^{1+s}}
\right ) \right ] \, .
\end{equation}
This expression can be rearranged by separating the odd and even
$s$ contributions and by some simplifications
\begin{equation}
f(x_v)=\frac {1}{1-x_v}\prod _{s=0}^\infty \frac
{(1-x_vq_v^{-1}p_v^{1+2s})(1-x_vq_vp_v^{2s})}{(1-x_vq_v^{-1}p_v^{2+2s})
(1-x_vq_vp_v^{1+2s})} \, .
\end{equation}
Using notation (\ref {infprod}) for infinite products, namely
\begin{equation}
(x;a)=\prod _{s=0}^{\infty}(1-xa^s) \, , \nonumber
\end{equation}
we can write down the compact expression
\begin{equation}
f(x_v)=\frac {1}{1-x_v} \frac
{(x_vq_v^{-1}p_v;p_v^2)(x_vq_v;p_v^2)}{(x_vq_v^{-1}p_v^2;
p_v^2)(x_vq_vp_v;p_v^2)} \, . \label{fDVA3}
\end{equation}
Now, we define a meromorphic function, the so-called structure
function of the DVA, as
\begin{equation}
Y(x_v)=\frac {f(x_v)}{f(x_v^{-1})}=-\frac {1}{x_v}\frac
{(x_vq_v^{-1}p_v;p_v^2)(x_vq_v;p_v^2)(x_v^{-1}q_v^{-1}p_v^2;
p_v^2)(x_v^{-1}q_vp_v;p_v^2)}{(x_v^{-1}q_v^{-1}p_v;p_v^2)(x_v^{-1}q_v;
p_v^2)(x_vq_v^{-1}p_v^2;p_v^2)(x_vq_vp_v;p_v^2)} \, . \label
{Yfunct}
\end{equation}
This function enters the exchange algebra
\begin{equation}
T(z)T(w)=Y(z/w)T(w)T(z) \, ,
\end{equation}
which is of course equivalent to the DVA.

\vspace {0.8cm}

\subsection{Breather factor $S_B(\theta)$ in Lukyanov's form}
We want to elaborate the DVA expression (\ref{DVAequ}, \ref{DVA1})
for $S_B(\theta)$ following an idea by Lukyanov (\cite {LUK}).
First, keeping in mind (\ref{DVAnota}), we collect the infinite
products of (\ref{DVA1}) within theta-functions of nome
$p_v={\mbox {exp}}\left (-\frac {4 \eta \pi}{{\bf K}'} \right )$:
\begin{equation}
S_B(\theta)=\frac {\theta _{11} \left ( -\frac {\eta \theta}{{\bf
K}{\bf K}'}+2i \frac {{\bf K}}{{\bf K}'}; \frac {4i \eta}{{\bf
K}'}\right ) \theta _{01} \left ( -\frac {\eta \theta}{{\bf K}{\bf
K}'}- 2i \frac {{\bf K}}{{\bf K}'}; \frac {4i \eta}{{\bf
K}'}\right )} {\theta _{11} \left ( -\frac {\eta \theta}{{\bf
K}{\bf K}'}-2i \frac {{\bf K}}{{\bf K}'}; \frac {4i \eta}{{\bf
K}'}\right ) \theta _{01} \left ( -\frac {\eta \theta}{{\bf K}{\bf
K}'}+ 2i \frac {{\bf K}}{{\bf K}'}; \frac {4i \eta}{{\bf
K}'}\right )} \, .
\end{equation}
Then, we use the modular transformation
\begin{equation}
\frac {\theta _{11}\left(\alpha; i\frac {K'}{K}\right)}{\theta
_{01}\left(\alpha ; i\frac {K'}{K}\right)}= i \frac {\theta
_{11}\left (-i \frac {K}{K'}\alpha; i\frac {K}{K'}\right )}
{\theta _{10}\left (-i \frac {K}{K'}\alpha; i\frac {K}{K'}\right
)} \, ,
\end{equation}
which connects theta-functions with nome $p={\mbox {exp}}\left
(-\pi \frac {K'}{K} \right )$ (on the l.h.s.) with theta-functions
with nome $p'={\mbox {exp}}\left (-\pi \frac {K}{K'} \right )$ (on
the r.h.s.). This allows us to re-express $S_B(\theta)$ as follows
\begin{equation}
S_B(\theta)=\frac {\theta _{11} \left ( \frac {i\theta}{4{\bf
K}}+\frac {{\bf K}}{2\eta}; \frac {i {\bf K}'}{4{\eta}}\right )
\theta _{10}\left ( \frac {i\theta}{4{\bf K}}-\frac {{\bf
K}}{2\eta}; \frac {i {\bf K}'}{4{\eta}}\right ) } {\theta _{11}
\left ( \frac {i\theta}{4{\bf K}}-\frac {{\bf K}}{2\eta}; \frac {i
{\bf K}'}{4{\eta}}\right ) \theta _{10}\left ( \frac
{i\theta}{4{\bf K}}+\frac {{\bf K}}{2\eta}; \frac {i {\bf
K}'}{4{\eta}}\right ) } \, ,
\end{equation}
which is (\ref {DVAluk}) of the main text. Upon comparing this
relation with (14) of \cite {LUK}, we must identify Lukyanov's
variables (on the left) and ours in this manner:
\begin{equation}
\beta _{L}=\frac {\pi \theta}{2{\bf K}} \, , \quad \xi _{L}=\frac
{{\bf K}}{\eta}-1 \, , \quad x_{L}= {\mbox {exp}}\left (-\frac {2
\eta \pi}{{\bf K}'} \right ) \, . \label {usandluk}
\end{equation}

\vspace {1cm}

\subsection{Breather factor $S_B(\theta)$ in Mussardo-Penati's form}
Here we want to show that Mussardo-Penati scattering factor
$S_{MP}(\beta)$ actually coincides with Lukyanov's one and
therefore with our $S_B(\theta)$. We start from expression (6) of
\cite{MP} for $S_{MP}(\beta)$ and simply re-formulate it by
expressing the Jacobian elliptic functions in terms of
theta-functions:
\begin{equation}
S_{MP}(\beta)=\frac {\theta _{11}\left (\frac {\beta -i\pi
a}{2i\pi};\frac {iT}{\pi} \right ) \theta _{10}\left (\frac {\beta
+i\pi a}{2i\pi};\frac {iT}{\pi}\right ) \theta _{01}\left (\frac
{\beta -i\pi a}{2i\pi};\frac {iT}{\pi}\right )\theta _{00}\left
(\frac {\beta +i\pi a}{2i\pi};\frac {iT}{\pi}\right )} {\theta
_{11}\left (\frac {\beta +i\pi a}{2i\pi};\frac {iT}{\pi}\right
)\theta _{10}\left (\frac {\beta -i\pi a}{2i\pi};\frac
{iT}{\pi}\right ) \theta _{01}\left (\frac {\beta +i\pi
a}{2i\pi};\frac {iT}{\pi}\right )\theta _{00}\left (\frac {\beta
-i\pi a}{2i\pi};\frac {iT}{\pi}\right )} \, .
\end{equation}
As stressed the elliptic nome of these theta-functions is
$e^{-T}$. Now, we use the simple identities (deriving from the
definitions),
\begin{eqnarray}
\theta _{11}(u;\tau)\theta _{01}(u;\tau)&=&\theta _{11}(u;\tau/2)
\frac
{(e^{2i\pi\tau};e^{2i\pi\tau})^2}{(e^{i\pi\tau};e^{i\pi\tau})}
e^{\frac
{i\pi \tau}{8}} \, , \\
\theta _{10}(u;\tau)\theta _{00}(u;\tau)&=&\theta _{10}(u;\tau/2)
\frac
{(e^{2i\pi\tau};e^{2i\pi\tau})^2}{(e^{i\pi\tau};e^{i\pi\tau})}
e^{\frac {i\pi \tau}{8}} \, ,
\end{eqnarray}
in order to express $S_{MP}(\beta)$ in the final form
\begin{equation}
S_{MP}(\beta)=\frac {\theta _{11}\left (\frac {\beta -i\pi
a}{2i\pi};\frac {iT}{2\pi}\right ) \theta _{10}\left (\frac {\beta
+i\pi a}{2i\pi};\frac {iT}{2\pi}\right )} {\theta _{11}\left
(\frac {\beta +i\pi a}{2i\pi};\frac {iT}{2\pi}\right ) \theta
_{10}\left (\frac {\beta -i\pi a}{2i\pi};\frac {iT}{2\pi}\right )}
\, .
\end{equation}
Indeed, we notice that this coincides with $S_B(\theta)$
(\ref{DVAluk}), provided we link Mussardo-Penati's variables (on
the l.h.s.) with ours (on the r.h.s.) according to
\begin{equation}
T_{MP}=\frac {\pi {\bf K}'} {2\eta} \, , \quad a_{MP}=\frac {{\bf
K}}{\eta }-1 \, , \quad \beta_{MP}=-\frac {\pi \theta}{2{\bf K}}
\, .
\end{equation}

\end{document}